%% file: dim6.tex
\documentclass[12pt,a4paper]{article}
\usepackage{epsf,epsfig,epstopdf,rotating,axodraw}
\begin{document}
\begin{flushright}
{SINP MSU 2013-2/885\\ 
. \\
September 2013}
\end{flushright}
\vspace{0.5cm}

\begin{center}
{\Large \bf Higgs boson signal at complete tree level in the SM extension by dimension-six operators
\hfill\\}
\end{center}
\vspace{0.5cm}

\begin{center}
{E.~Boos$^{1}$, V.~Bunichev$^{1}$, M.~Dubinin$^{1}$, Y.~Kurihara$^2$\\
\hfill\\
{\small \it $^1$Institute of Nuclear Physics, Moscow State University}\\
{\small \it 119991, Moscow, Russia} \\
 {\small \it $^2$High Energy Accelerators Research Organization (KEK), Tsukuba, Ibaraki-ken, Japan} }
\end{center}

\vspace{1.0cm}
\begin{center}
{\bf Abstract}
\end{center}
\begin{quote}
{\small Deviations from the standard Higgs sector generated by some new (nonstandard) physics at an energy scale $\Lambda$ could be described by an effective $SU(3)_c \times SU(2)_L \times U(1)$ invariant non-renormalizable Lagrangian terms of dimension six. The set of dimension-six operators involving the Higgs field is chosen in such a way that the form of gauge bosons kinetic terms remains untouched, preserving all high-precision electroweak constraints. A systematic study of effects in various Higgs boson production channels ($\gamma \gamma$, $ZZ$, $WW$, $b \bar b$, $\tau \bar \tau$ ) caused by effective operators is carried out beyond the {\it production $\times$ decay} approximation (or infinitely small width approximation). Statistical methods are used to establish consistency of the standard Higgs sector with the available LHC data. A global fit in the two-parametric anomalous coupling space indicating to possible deviations from the standard Higgs-fermion and Higgs-gauge boson couplings is performed, using post-Moriond 2012 data and more precise LC 2013 data. We find that the standard Higgs sector is consistent with the current CMS and ATLAS experimental results both in the infinitely small width approximation and the calculation with complete gauge invariant sets of diagrams. However, visible difference of the exclusion contours is found for some combinations of production and decay channels, although minor for the global fits for all possible channels. Updates of the signal strength and the signal strength error reported at LC 2013 result in a significant improvements of the allowed regions in the anomalous coupling space, which are recalculated also at complete tree level.}
\end{quote}   

\newpage

\section{Introduction}

Discovery of a Higgs-like signal at the LHC \cite{signal} provides a possibility to accomplish the Standard Model (SM) scheme, which is however considered as an effective theory at the energy scale $v=$246 GeV rather than a self-contained field theory model. Physical observables up to energies of the order of 'new physics' scale $\Lambda$ are described by an effective Lagrangian which can be written as an expansion in the inverse powers of 
$\Lambda$. It is assumed usually that electroweak symmetry breaking scale $v$ is disconnected with the scale of new physics $\Lambda$, so the effective Lagrangian terms are invariant with respect to the gauge group $SU(3)_c\times SU(2)_L \times U(1)_Y$. Effective operators, first introduced in connection with a hypothetical baryon number violation and the four-fermion contact interactions \cite{first}, then were used \cite{following} to consider flavor-changing neutral currents, extended technicolor, composite models and other BSM extensions. In the following we are using results of a systematic study \cite{buchmuller}, where sector-by-sector extensions of the SM by dimension 5 and dimension 6 effective operators can be found. Improved classification of anomalous terms, where some redundant dimension-six operators were excluded, can be found in \cite{grzadkowski}. An equivalent basis, which isolates in a more convenient manner the operators essential for the decays $H\to \gamma \gamma$, $\gamma Z$ has been elaborated in \cite{contino_basis} including higher order corrections. Insofar as the effective Lagrangian terms include classical fields, equations of motion can be used for simplifications\footnote{Correspondingly, a set of Schwinger-Dyson equations can be used at the quantum level.} in both the scalar-fermion and scalar-gauge boson sectors, with result for the following set of dimension 6 operators
\begin{equation}
\label{setb}
\end{equation}
\vspace{-1.5cm}
\begin{itemize}
\item{{\it scalar-gauge boson sector\\}
\begin{tabular}{cc}
$O_{\Phi G} = \frac{1}{2}(\Phi^\dagger\Phi-\frac{v^2}{2})G^a_{\mu\nu}G^{a\mu\nu}$ & $O_{\Phi G} = \frac{1}{2}(\Phi^\dagger\Phi-\frac{v^2}{2})G^a_{\mu\nu}\tilde{G}^{a\mu\nu}$ \\
$O_{\Phi B} = \frac{1}{2}(\Phi^\dagger\Phi-\frac{v^2}{2})B_{\mu\nu}B^{\mu\nu}$ & $O_{\Phi B} = \frac{1}{2}(\Phi^\dagger\Phi-\frac{v^2}{2})B_{\mu\nu}\tilde{B}^{\mu\nu}$ \\ 
$O_{\Phi W} = \frac{1}{2}(\Phi^\dagger\Phi-\frac{v^2}{2})W^i_{\mu\nu}W^{i\mu\nu}$ & $O_{\Phi W} = \frac{1}{2}(\Phi^\dagger\Phi-\frac{v^2}{2})W^i_{\mu\nu}\tilde{W}^{i\mu\nu}$ \\
$O^{(1)}_{\Phi} = (\Phi^\dagger\Phi-\frac{v^2}{2})D_{\mu}\Phi^\dagger D^{\mu}\Phi$ & 
\end{tabular}}
\begin{equation}
\vspace{-1.3cm}
\label{setf}
\end{equation}
\item{{\it scalar-fermion sector\\}
$O_{t\Phi} = (\Phi^\dagger\Phi-\frac{v^2}{2})(\bar{Q_L}\Phi^c t_R)$ \\
$O_{b\Phi} = (\Phi^\dagger\Phi-\frac{v^2}{2})(\bar{Q_L}\Phi b_R)$ \\
$O_{\tau \Phi} = (\Phi^\dagger\Phi-\frac{v^2}{2})(\bar{L_L}\Phi \tau_R)$
}
\end{itemize}
where dual tensor $\tilde{F}_{\mu \nu}=\epsilon_{\mu \nu \gamma \delta} F_{\gamma \delta}$.
Deviations from the SM are defined by the effective Lagrangian
\begin{equation}
 L_{eff}^{(6)} \, = \, \frac{1}{\Lambda^2} \sum_{k=V,F} C_{k \Phi} O_{k \Phi} 
\end{equation}
where the anomalous couplings $C$ modify the SM Higgs boson couplings to the vector bosons and to the fermions. 

The subtraction of $v^2/2$ leaves out undesirable mixing in the gauge field kinetic terms. Such operator basis was considered in \cite{passarino}. Reduced set of five operators for anomalous Higgs couplings to gauge bosons only was analysed in \cite{eboli}, where additional operators containing covariant derivatives of the scalar doublet, $O_W=D_\mu \Phi^\dagger D_\nu \Phi \, W^{\mu \nu}$ and $O_B=D_\mu \Phi D_\nu \Phi^\dagger B^{\mu \nu}$, modifying the triple gauge-boson couplings, were either accepted or rejected in the two independent scenarios. The operator $O_{BW}=\Phi^\dagger \tau^a \Phi B_{\mu \nu} W^{\mu \nu}$, contributing to the $W^3-B$ mixing of $SU(2)$ eigenstates at the tree level is strongly constrained by the electroweak data. 
Mixing terms with derivatives of vector fields would imminently shift gauge boson masses which are severely constrained by the electroweak precision data.  The operator $(\Phi^+ \Phi)^3$ (denoted also by $O_{\Phi}^{(3)}$) which shifts the minimum of the Higgs potential and the Higgs boson mass is not introduced in our analysis.

In the general case when all possible dimension six contributions are accounted for, the effective Higgs-boson and Higgs-fermion couplings are related to the coefficients $C_{\Phi k}$ in front of the operators $O_{\Phi G}$, .... $O_{\tau \Phi}$ in a rather untrivial way, since a number of different coefficients $C_{\Phi k}$ mix while contributing to a single effective Higgs-boson or Higgs-fermion coupling. For the sake of distinctness we restrict the general multidimensional anomalous coupling space to the two-dimensional space\footnote{Analysis of the complete set of operators is in progress and will be presented separately.}, where the Higgs-boson and the Higgs-fermion couplings are rescaled by independent parameters $c_V$ and $c_F$. Such reduction is meaningful also to avoid modifications of the Lorentz structure for a vertex. Nonzero anomalous couplings $C_{\Phi b}$ and $C_{\Phi W}$, for instance, lead to modifications of the tensor structure for $H W^+ W^-$ and $HZZ$ vertices (see Section 2 and Table 2, more details can be found in \cite{tensor}). As a result, the phase space distributions could be substantially modified \cite{distributions} in comparison with the SM, making questionable the experimental interpretation of the signal reconstruction which is based specifically on the SM. Linear rescailing of the Higgs-boson and the Higgs-fermion couplings ($c_V$,$c_F$) (sometimes denoted by $k_v$ and $k_f$) is a common feature of a majority of existing analyses (see more comprehensive list \cite{fullset}) which refer to different theoretical backgrounds. Particular parametrization of the Higgs couplings of the form $k_f=\sqrt{2}(m_f/M)^{1+\epsilon}$, $k_v=2(m_V^{2(1+\epsilon)}/M^{1+2\epsilon})$ (specific to the genuine Englert-Brout-Higgs spontaneous symmetry breaking mechanism; the limiting case of the SM is $\epsilon=0$, $M=v$) has been analysed in \cite{ellis_you}. One can distinguish a group of approaches where the fundamental scalar field is not a component of $SU(2)$ doublet \cite{ellis_you_0,silh}. Anomalous operators of dimension five form the corresponding effective operator basis in the framework of a nonlinear realization of the $SU(2)_L \times U(1)$ symmetry by means of an effective chiral Lagrangian. In such models the Goldstone bosons $\pi^a$ are introduced in the form of a field $\Sigma(x)=exp(i \tau^a \pi_a/v)$, which transforms linearly under the group $SU(2)_L \times SU(2)_R$. Parameter $v$ is not a vacuum expectation value associated with a minimum of some potential,  the Higgs field is an additional scalar singlet under the gauge group transformations. An effective Lagrangian in the form of expansion in the powers of $h/v$ can be found in \cite{ellis_you}-\cite{silh_lgrng}. The effective parameters $a,b,...$ in front of various powers of $h/v$ in the expansion define the values of $H$ couplings to gauge bosons, fermions and $H$ self-interaction. The leading operators appearing in the expansion in the inverse powers of the cutoff scale $\Lambda$ have the dimension five \cite{silh}
\begin{eqnarray*}
L_{eff}^{(5)} = -\frac{c_g g^2_s}{2\Lambda} h G^A_{\mu \nu} G^{A \mu \nu} - \frac{c_W g^2_2}{2\Lambda} h W^a_{\mu \nu} W^{a \mu \nu} -\frac{c_B g^2_1}{2\Lambda} h B_{\mu \nu} B^{ \mu \nu}
\end{eqnarray*}        
and are enhanced by the factor $\Lambda/v$ in comparison with the effective dim-6 operators. In the minimal composite pseudo-Goldstone boson scenario the Higgs-boson and the Higgs-fermion anomalous couplings are identically rescaled, but it is not the case for a nonminimal compositeness, when some higher-order chiral symmetry is broken down to the symmetry of the standard Higgs sector. A scenario where light composite Higgs boson, emerging from a strongly interacting sector as a pseudo-Goldstone boson, causes electroweak symmetry breaking has been analysed in details \cite{ellis_you}-\cite{fit1} in connection with LHC data. The Higgs-fermion effective terms lead to a number of observable consequences for vector boson scattering and enhanced double Higgs boson production \cite{silh_lgrng,silheffects}.

Recent updates of CMS and ATLAS results in $\gamma \gamma$ and $ZZ$,$WW$ channels \cite{cms_pas} for the standard model Higgs boson allow to improve previous considerations, analyzing consistency of an experimental data to expectations for the SM Higgs boson production. 
Note that these analyses are based on a phenomenological parametrisation \cite{hxswg} specific to production $\times$ decay approximation, when the Higgs boson width is infinitely small, the Breit-Wigner propagator is replaced by delta function so the signal cross section for the channel $ii\to H\to ff$ is $\sigma_{ii}(ii\to H) \times \Gamma_{ff}/\Gamma_{tot}$. While "dressing" cross-sections $\sigma_{ii}$ and decay widths $\Gamma_{ff}$ by the scale factors $k_{i,f}$, factorizable deviations from the SM are introduced. For example, in the channel $gg\to H\to \gamma \gamma$ anomalous factor has a simple form $k^2_g k^2_\gamma/k^2_H$. The factors $k_g$ and $k_\gamma$ are independent parameters, i.e. vector boson and fermion loops are not resolved.      

In the following we are going to analyse the LHC results in the framework of the SM extension by the dimension six operators and clear up the consistency of data and the consequences of the SM for Higgs-fermion and Higgs-gauge boson couplings\footnote{ATLAS and CMS results reported in the beginning of 2013 (Recontres de Moriond, \cite{moriond})  were substantially improved in May 2013 (European LC Workshop).
ATLAS results in the $\gamma \gamma$ channel: significance 7.4$\sigma$, $\mu=$1.65+0.34-0.30  (2.3$\sigma$ above the expected for SM), $m_H=$126.8$\pm$0.2(stat)$\pm$0.7(syst) GeV; in the $ZZ$ channel:  significance 6.6$\sigma$, $\mu$=1.70+0.5-0.4, $m_H=$124.3+0.6-0.5(stat)+0.5-0.3(syst) GeV; 
CMS results in the $\gamma \gamma$ channel: significance 7$\sigma$, $\mu=$1.55$\pm$0.5 consistent with $m_H\sim$125 GeV; in the $ZZ$ channel:  significance 6.6$\sigma$, $\mu$=0.91+0.30-0.24, $m_H=$125.8$\pm$0.5(stat)$\pm$0.2(syst) GeV. Improvements in May 2013  can be found in section 4.}.
The paper is organized as follows. In section 2 convenient normalization of effective vertices in the dimension-six operator basis is defined. Section 3 contains statistical analysis of Higgs production data. Results of our computations are summarized in section 4.

\section{Normalization of effective vertices}

A set of P-conserving operators, Eqs.(\ref{setb}),(\ref{setf}), leads to the set of Feynman rules listed in Table 1.

\begin{table}
\begin{center}
\begin{tabular}{|l|l|l|} \hline
~~~~~Effective operators & Triple vertices & ~~~~~~~~~~~~Feynman rules\\ \hline
\phantom{-} & \phantom{-} & \phantom{-} 
\\[1mm]
$O_{t\Phi} = (\Phi^\dagger\Phi-\frac{v^2}{2})(-\lambda_t)(\bar{Q_L}\Phi^c t_R)$ &
$\bar{t}$ \phantom{-} $t$ \phantom{-} ${H}_{}$ \phantom{-}  &
        $-M_t\cdot \frac{v}{\Lambda^2}\cdot C_{t\Phi}$
\\[1mm]
$O_{b\Phi} = (\Phi^\dagger\Phi-\frac{v^2}{2})(-\lambda_b)(\bar{Q_L}\Phi b_R)$ &
$\bar{b}$ \phantom{-} $b$ \phantom{-} ${H}_{}$ \phantom{-}  &
        $-M_b\cdot \frac{v}{\Lambda^2}\cdot C_{b\Phi}$
\\[1mm]
$O_{\tau \Phi} = (\Phi^\dagger\Phi-\frac{v^2}{2})(-\lambda_{\tau})(\bar{L_L}\Phi
\tau_R)$ &
$\bar{\tau}$ \phantom{-} $\tau$ \phantom{-} ${H}_{}$ \phantom{-}  &
        $-M_{\tau}\cdot \frac{v}{\Lambda^2}\cdot C_{\tau \Phi}$
\\[1mm]
\phantom{-} & \phantom{-} & \phantom{-} 
\\[1mm]
$O_{\Phi G} = \frac{1}{2}(\Phi^\dagger\Phi-\frac{v^2}{2})G^a_{\mu\nu}G^{a\mu\nu}$ &
${G}_{\mu}$ \phantom{-} ${G}_{\nu}$ \phantom{-} ${H}_{}$ \phantom{-}  &
        $-2\cdot \frac{v}{\Lambda^2}\cdot C_{\Phi G}\cdot \big(g^{\mu \nu} p_1 p_2 
-p_1^\nu p_2^\mu \big)$
\\[1mm]
\phantom{-} & \phantom{-} & \phantom{-} 
\\[1mm]
$O_{\Phi B} = \frac{1}{2}(\Phi^\dagger\Phi-\frac{v^2}{2})B_{\mu\nu}B^{\mu\nu}$ &
${A}_{\mu }$ \phantom{-} ${A}_{\nu }$ \phantom{-} ${H}_{}$ \phantom{-}  &
        $-2\cdot c_w^2\cdot \frac{v}{\Lambda^2}\cdot C_{\Phi B}\cdot \big(g^{\mu \nu} p_1
p_2  -p_1^\nu p_2^\mu \big)$
\\[1mm]
\phantom{-} &
${A}_{\mu }$ \phantom{-} ${Z}_{\nu}$ \phantom{-} ${H}$ \phantom{-}  &
        $+2\cdot c_w\cdot s_w\cdot \frac{v}{\Lambda^2}\cdot C_{\Phi B}\cdot \big(g^{\mu
\nu} p_1 p_2  -p_1^\nu p_2^\mu \big)$
\\[1mm]
\phantom{-} &
${Z}_{\mu}$ \phantom{-} ${Z}_{\nu }$ \phantom{-} ${H}$ \phantom{-}  &
        $-2\cdot s_w^2\cdot \frac{v}{\Lambda^2}\cdot C_{\Phi B}\cdot \big(g^{\mu \nu} p_1
p_2  -p_1^\nu p_2^\mu \big)$
\\[1mm]
\phantom{-} & \phantom{-} & \phantom{-} 
\\[1mm]
$O_{\Phi W} = \frac{1}{2}(\Phi^\dagger\Phi-\frac{v^2}{2})W^i_{\mu\nu}W^{i\mu\nu}$ &
${A}_{\mu }$ \phantom{-} ${A}_{\nu }$ \phantom{-} ${H}_{}$ \phantom{-}  &
        $-2\cdot s_w^2\cdot \frac{v}{\Lambda^2}\cdot C_{\Phi W}\cdot \big(g^{\mu \nu} p_1
p_2  -p_1^\nu p_2^\mu \big)$
\\[1mm]
\phantom{-} &
${A}_{\mu }$ \phantom{-} ${Z}_{\nu}$ \phantom{-} ${H}$ \phantom{-}  &
        $-2\cdot c_w\cdot s_w\cdot \frac{v}{\Lambda^2}\cdot C_{\Phi W}\cdot \big(g^{\mu
\nu} p_1 p_2  -p_1^\nu p_2^\mu \big)$
\\[1mm]
\phantom{-} &
${Z}_{\mu }$ \phantom{-} ${Z}_{\nu }$ \phantom{-} ${H}_{}$ \phantom{-}  &
        $-2\cdot c_w^2\cdot \frac{v}{\Lambda^2}\cdot C_{\Phi W}\cdot \big(g^{\mu \nu} p_1
p_2  -p_1^\nu p_2^\mu \big)$
\\[1mm]
\phantom{-} &
$W^+_{\mu }$ $W^-_{\nu }$ ${H}$   &
        $-2\cdot \frac{v}{\Lambda^2}\cdot C_{\Phi W}\cdot \big(g^{\mu \nu} p_1 p_2 
-p_1^\nu p_2^\mu \big)$
\\[1mm]
\phantom{-} & \phantom{-} & \phantom{-} 
\\[1mm]
$O^{(1)}_{\Phi} = (\Phi^\dagger\Phi-\frac{v^2}{2})D_{\mu}\Phi^\dagger D^{\mu}\Phi$ &
$W^+_{\mu }$ $W^-_{\nu }$  ${H}_{}$   &
        $M_W^2\cdot \frac{v}{\Lambda^2}\cdot C^{(1)}_{\Phi}\cdot g^{\mu \nu} $
\\[1mm]
\phantom{-} &
${Z}_{\mu }$ \phantom{-} ${Z}_{\nu }$ \phantom{-} ${H}_{}$ \phantom{-}  &
        $M_Z^2\cdot \frac{v}{\Lambda^2}\cdot C^{(1)}_{\Phi}\cdot g^{\mu \nu} $
\\[1mm]
 \hline
\end{tabular}
\end{center}
\caption{Effective triple vertices in the Buchmueller-Wyler operator basis. Anomalous couplings $C$ (Wilson coefficients) are multiplicative factors in front of $O$.}
\end{table}

As already mentioned in the Introduction, the following analysis will be focused on the Higgs-fermion and the Higgs-gauge boson anomalous couplings $C_{t \Phi}$, $C_{b \Phi}$, $C_{\tau \Phi}$ and $C^{(1)}_\Phi$, $C_{\Phi G}$, which conserve the SM Lorentz structure of vertices. It is convenient to use a parametrisation which gives explicitly the SM one-loop contributions for the Higgs decays at some point of the anomalous couplings parameter space. If the effective Lagrangians for $H\to \gamma \gamma$ and $H\to gg$ are
\begin{equation}
L^{eff}_{\gamma \gamma H}=\frac{\lambda_{\gamma \gamma H}}{4} F_{\mu \nu} F^{\mu \nu} H, \quad
L^{eff}_{ggH}=\frac{\lambda_{g g H}}{4} G^a_{\mu \nu} G^{a \mu \nu} H
\end{equation}
then the effective vertices 
\begin{equation}
\Gamma^{\mu \nu}(p_1,p_2)_{\gamma / g}=-\lambda_{\gamma \gamma H / ggH}(g^{\mu \nu} p_1 p_2-p^\nu_1 p^\mu_2) 
\end{equation}
where $\lambda_{\gamma \gamma H / ggH}$ are defined by the one-loop integrals. The dominant fermionic contribution of the top-quark loop leads to well-known effective Lagrangians (see details in the survey \cite{05djouadi}) ($\sqrt{G_F \sqrt{2}}=1/v$)
\begin{equation}
L^{eff}_{\gamma \gamma H} = \frac{2\alpha}{9\pi v} F_{\mu \nu} F^{\mu \nu} H, \quad
L^{eff}_{ggH}=-\frac{\alpha_s}{12\pi v} G^a_{\mu \nu} G^{a \mu \nu} H
\end{equation} 
for the case of rather small $m^2_H/4m^2_{top}$ which is valid satisfactory\footnote{In the numerical computations well-known formulae including $m^2_H/4m^2_{top}$ terms were used.} for $m_H=$126 GeV. So for the top one-loop induced couplings
\begin{equation}
\lambda^t_{\gamma \gamma H}= \frac{8\alpha}{9\pi v}, \quad \lambda^t_{ggH}=-\frac{\alpha_s}{3\pi v}
\label{toploop}
\end{equation}
The contribution of $W$ is known for a long time \cite{76ellis}
\begin{equation}
\lambda^W_{\gamma \gamma H} = -\frac{7\alpha}{2\pi v}
\label{wloop}
\end{equation}
and the one-loop induced decay widths are \cite{05djouadi}
\begin{eqnarray}
\Gamma(H\to \gamma \gamma) &=& \frac{\alpha^2 G_F m^3_H}{128\pi^3 \sqrt{2}} \left| 3 \left( \frac{2}{3} \right)^2 \frac{4}{3} -7\right|^2 = (\frac{47}{9})^2 \frac{\alpha^2 G_F m^3_H}{128\pi^3 \sqrt{2}}, \\
\Gamma(H\to gg) &=& \frac{1}{36} \frac{\alpha^2 G_F m^3_H}{\pi^3 \sqrt{2}}. 
\end{eqnarray}
It is convenient to introduce the effective parameters
\begin{eqnarray*}
c_F&=& 1 + C_{t\Phi}\cdot\frac{v^2}{\Lambda^2}\\
c_V&=& 1+\frac{v^2}{2\Lambda^2}\cdot C^{(1)}_{\Phi}\\
c_G&=& c_F+\frac{6\pi}{\alpha_s}\cdot C_{\Phi G}\cdot \frac{v^2}{\Lambda^2}\\
c_{\gamma}&=& \frac{63 c_F-16 c_V}{47} +
\frac{9\pi}{4\alpha}\cdot(c_w^2\cdot C_{\Phi B} + s_w^2\cdot C_{\Phi
W})\cdot\frac{v^2}{\Lambda^2}\\
c_Z&=& (s_w^2\cdot C_{\Phi B} + c_w^2\cdot C_{\Phi W})\cdot\frac{v^2}{\Lambda^2}\\
c_W&=& C_{\Phi W}\cdot\frac{v^2}{\Lambda^2}
\end{eqnarray*}
such that at the leading order the SM limit with the one-loop induced $H\to \gamma \gamma$ and $h\to gg$ channels is clearly seen.  A compact set of Feynman rules for the triple vertices to be used in the following analyses is presented in Table 2. In order to take into account the NLO corrections, the normalization of $ggH$ and $\gamma \gamma H$ vertices was changed using the output of HDECAY code \cite{hdecay}, where higher-order QCD and leading electroweak corrections from different sources have been incorporated. For example, the effective coupling constants $\lambda_{\gamma \gamma H}$ and $\lambda_{\gamma \gamma Z}$ can be found using partial widths 
\begin{equation}
\lambda_{\gamma \gamma H}=8
\sqrt{\frac{\pi}{m^3_H} \Gamma^{tot} Br(H \to \gamma \gamma)}, \quad
\lambda_{\gamma Z H}=8
\sqrt{\frac{\pi m^3_H}{2(m^2_H-m^2_Z)^3}\Gamma^{tot} Br(H \to \gamma Z)}
\end{equation}
Such normalization reproduces the SM limit at $c_i=$1, $i=F,V,G,\gamma$, $c_Z=c_W=0$.  The one-loop vertices are "resolved" at the leading order. For example, destructive interference between the top and W loops, see Eqs.(\ref{toploop}) and (\ref{wloop}), leads to an enhancement in the $\gamma \gamma$ channel at negative $c_F$, where an extensive region compatible with the data appears (see Section 4). However, NLO corrections from anomalous dim-6 terms inside the loops are not accounted for.
\begin{table}
\begin{center}
\begin{tabular}{|l|l|} \hline
Triple vertices & Feynman rules \\ \hline
$\bar{t}$ \phantom{-} $t$ \phantom{-} ${H}_{}$ \phantom{-}  &
        $-\frac{M_t}{v}\cdot c_F$\\[2mm]
$\bar{b}$ \phantom{-} $b$ \phantom{-} ${H}_{}$ \phantom{-}  &
        $-\frac{M_b}{v}\cdot c_F$\\[2mm]
$\bar{\tau}$ \phantom{-} $\tau$ \phantom{-} ${H}_{}$ \phantom{-}  &
        $-\frac{M_\tau}{v}\cdot c_F$\\[2mm]
${G}_{\mu}$ \phantom{-} ${G}_{\nu}$ \phantom{-} ${H}_{}$ \phantom{-}  &
        $-\frac{2}{v}\cdot \frac{\alpha_s}{6\pi}\cdot c_G\cdot \big(g^{\mu \nu} p_1 p_2 
-p_1^\nu p_2^\mu \big)$\\[2mm]
${A}_{\mu }$ \phantom{-} ${A}_{\nu }$ \phantom{-} ${H}_{}$ \phantom{-}  &
        $-\frac{2}{v}\cdot \frac{4\alpha}{9\pi}\cdot c_{\gamma}\cdot \big(g^{\mu \nu} p_1
p_2  -p_1^\nu p_2^\mu \big)$\\[2mm]
${A}_{\mu }$ \phantom{-} ${Z}_{\nu}$ \phantom{-} ${H}$ \phantom{-}  &
        $+2\cdot c_w\cdot s_w\cdot (C_{\Phi B} - C_{\Phi W})\cdot\frac{v}{\Lambda^2}
\big(g^{\mu \nu} p_1 p_2  -p_1^\nu p_2^\mu \big)$\\[2mm]
${Z}_{\mu }$ \phantom{-} ${Z}_{\nu }$ \phantom{-} ${H}_{}$ \phantom{-}  &
        $+\frac{2}{v}\cdot\left[M_Z^2\cdot c_V\cdot g^{\mu\nu} - c_Z\cdot \big(g^{\mu \nu}
p_1 p_2  -p_1^\nu p_2^\mu \big)\right]$\\[2mm]
$W^+_{\mu }$ \phantom{-} $W^-_{\nu }$ \phantom{-} ${H}_{}$ \phantom{-} &
$+\frac{2}{v}\cdot\left[M_W^2\cdot c_V\cdot g^{\mu\nu} - c_W\cdot \big(g^{\mu \nu} p_1
p_2  -p_1^\nu p_2^\mu \big)\right]$\\[2mm]
 \hline
\end{tabular}
\end{center}
\caption{Triple vertices in the Buchmueller-Wyler operator basis. The SM limit with the one-loop induced vertices 
$H\to \gamma \gamma$ and $h\to gg$ is achieved at $c_V=c_F=c_G=c_{\gamma}=$1, $c_Z=c_W=C_{\Phi B}=C_{\Phi W}=$0.}
\end{table}

\section{Signal strength and exclusion contours in the space of anomalous parameters}

The method of exclusion contours reconstruction in the relevant anomalous parameter space which we are using is similar to the method developed in \cite{fit1, fit2}. Available experimental data provides the signal strength
\begin{equation}
\mu_i = \frac{ [\sum^{}_j \sigma_{j \to h} Br(h\to i)]_{obs} }
             { [\sum^{}_j \sigma_{j \to h} Br(h\to i)]_{SM} }
\end{equation}
where $i$ is the number of Higgs boson decay channel and $j$ is the number of Higgs production process for a given final state. Best fit value of a signal strength can be expressed using the observed number of signal events $N_{obs}$, the number of background events $N_{backgr}$ and the number of signal events calculated in the SM $N^{SM}_{signal}$
\begin{equation}
{\hat \mu}_i = \frac{ N_{obs,i}-N_{backgr,i}}{N^{SM}_{signal,i}}
\label{ss}
\end{equation}
The global $\chi^2$ is defined as
\begin{equation}
\chi^2(\mu_i)=\sum^{N_{ch}}_{i} \frac{(\mu_i-{\hat \mu}_i)^2}{\sigma^2_i}
\end{equation}
for the number of production channels $N_{ch}$. Theoretical predictions for $\sigma_{j\to h}$ and related errors can be found on the LHC Higgs Cross Sections WG webpage \cite{lhchwg}. Minimization of $\chi^2\to \chi^2_{min}$ gives us the $1\sigma$, $2\sigma$ and $3\sigma$ regions $\chi^2= \chi^2_{min}+\Delta \chi^2$ where $\Delta \chi^2$ is defined by cumulative distribution function. Assuming that the signal strengths of various channels have Gaussian distributions with the probability density functions having the expected values ${\hat \mu}_i$ and the dispersions ($1\sigma$ deviations) $\sigma_i$ normalized to one, combined probability density function (pdf) for a number of production channels can be found by multiplication of pdf's for individual channels. Combined probability density 
function is also Gaussian characterized by $\mu_c$ and $\sigma_c$
\begin{equation}
\frac{1}{\sigma^2_c}=\sum^{N_{ch}}_i \frac{1}{\sigma^2_i}, \hskip 1cm 
\frac{{\hat \mu}_c}{\sigma^2_c}=\sum^{N_{ch}}_i \frac{{\hat \mu}_i}{\sigma^2_i}
\end{equation}
which allows to determine, for example, 95\% CL exclusion upper $\mu_U$ and lower $\mu_L$ limits on the signal strength parameter integrating the combined pdf from ${\hat \mu}$ to $\mu_U$ and from $\mu_L$ to ${\hat \mu}$, respectively, then equating result to 0.95/2. Possible negative values of the lower limit for the signal strength at a small luminosities allow to determine only $\mu_U$ by integrating probability density function from $0$ to $\mu_U$ and equating to 0.95. 

If the SM is fully adequate, the values of $\mu_i$ are as close to one as allowed by experimental errors. In the framework of the SM extension by dim-6 effective operators the values of $\mu_i$ obviously may depart from one for individual channels, so it is convenient to normalize the signal strengths (\ref{ss}) to the expectation values in the given SM extension $N_{signal,i,c_F,c_V,c_W...}$ rather than the SM expectation $N^{SM}_{signal,i}$, which does not depend on the anomalous parameters $c_F,c_V,c_W,...$. In this case the combined signal strength with expectation 1 can be introduced again if the exclusion bounds ${\hat \mu}_i \pm \sigma_i$ are renormalized by a factor $N^{SM}_{signal,i}/N_{signal,i,c_F,c_V,c_W...}$. While the experimental signal strength error is provided by ATLAS and CMS collaborations, the theoretical signal strength error is evaluated using numbers on the web page of LHC Higgs Cross Sections WG \cite{lhchwg}. 

In the following fits we are using the signal strength calculated at $m_H=$125 GeV. At the first stage a two-dimensional fit $\chi^2(c_V,c_F)$ has been performed, the anomalous couplings $C_{\Phi B}$ and $C_{\Phi W}$ (see Section 2) are taken to be zero, so for the SM case $(c_V,c_F)=(1,1)$, $c_G=c_\gamma=1$ and $c_Z=c_W=$0. Calculation of the $\Delta \chi^2$ for the best fit defines a {\it given number $\times$ CL} contours corresponding to the departure of the SM point $(1,1)$ from the best fit point in the $(c_V,c_F)$ parameter plane. Following \cite{fit1} the contours in all figures correspond to 65\%, 90\% and 99\% best fit CL regions with
$\Delta \chi^2$ less than 2.10, 4.61 and 9.21, respectively. 

\section{Beyond the infinitely small width approximation}

\begin{figure}[h]
\begin{center}
\input{ZZ.tex}
\end{center}
\caption{\label{zz} Signal diagrams for the process
$gg \to ZZ \to \mu^+ \mu^- \mu^+ \mu^-$}
\end{figure}
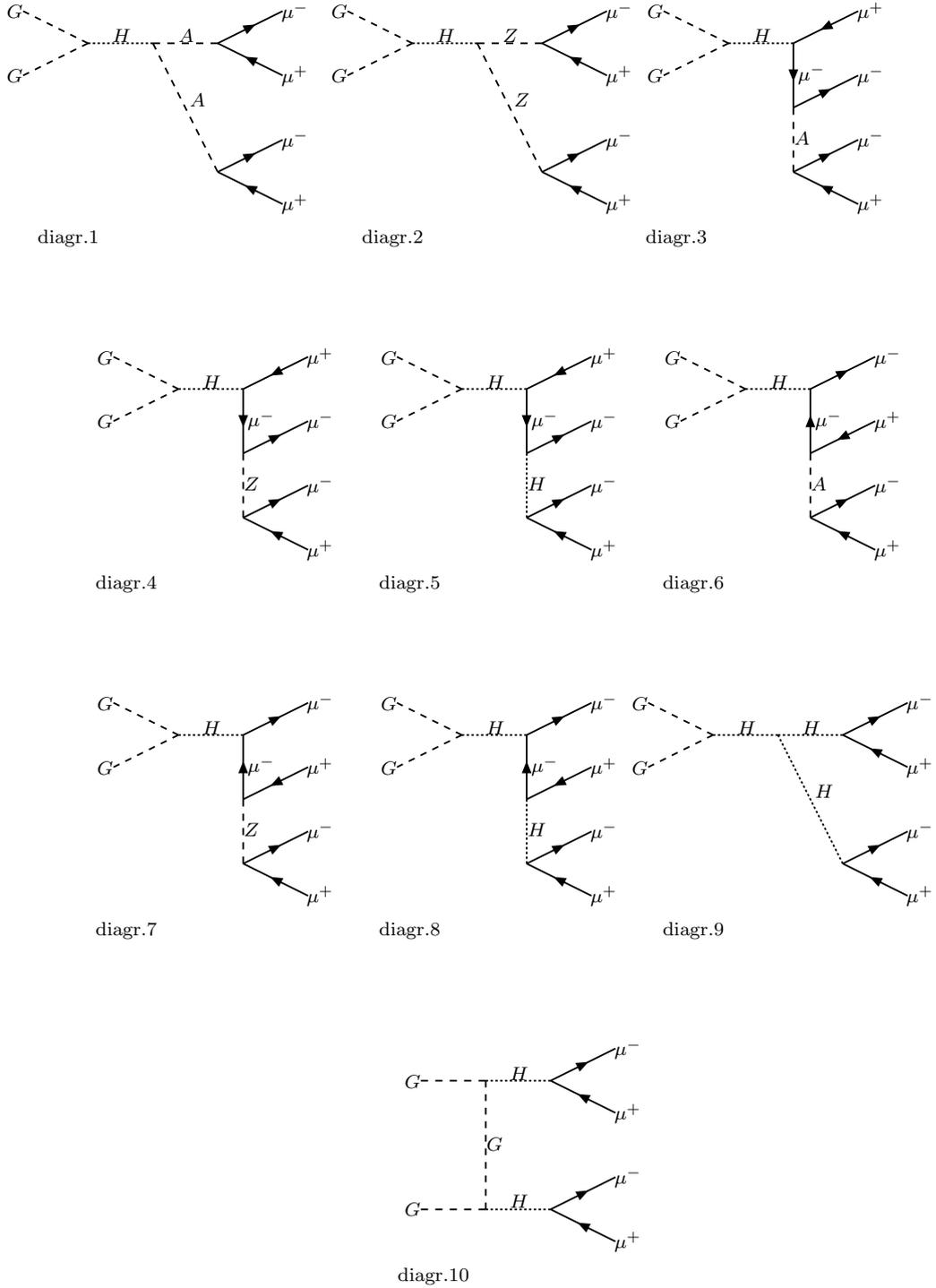

Calculations of complete gauge invariant sets of diagrams, although complicated and CPU time consuming, are more precise than production $\times$ decay approximation. They take into account
\begin{itemize}
\item{untrivial interference between signal diagrams. For example, the four-lepton final states $l^+ l^- l^+ l^-$ and/or $\nu_l \nu_l l^+ l^-$ are produced through $H\to Z^* Z^*$, $H\to W^* W^*$ and $\gamma \gamma$ intermediate states (see Figures \ref{ww} and \ref{zz}) with untrivial interferences, not accounted for in the production $\times$ decay approximation.}
\item{untrivial interference between the signal and the irreducible background diagrams. Although very small for narrow width resonances of the order of a few MeV, in the meaningful regions of the anomalous coupling space the anomalous Higgs boson width differs by around one order of magnitude from the SM total width. Numerical results for complete tree level sets are in some cases sensitive to Breit-Wigner propagators, especially when strong gauge cancellations between diagrams take place for a given Higgs production channel}
\item{lepton and jet distributions, calculated at complete tree level, are based on correct kinematics, oftenly not available for the production $\times$ decay approximation. Correct distributions are important in real experimental environment, when detector acceptances must be accounted for.}
\end{itemize}
Some theoretical issues and numerical examples in this connection were analysed for LEP2 physics \cite{LEP2}. 

A number of exclusion contours were reconstructed using the statistical approach described in Section 3.
The following Higgs production processes have been calculated 
\begin{itemize}
\item{{\it for the $\gamma \gamma$ event signature:} gluon fusion $gg\to \gamma \gamma$, vector boson fusion (VBF) $q q \to qq \gamma \gamma$, associated production with vector bosons $qq\to W \gamma \gamma$, $qq \to Z \gamma \gamma$ and the top-antitop quark pair $qq \to t \bar t \gamma \gamma$. Contributions to the Higgs boson production rate involving Higgs couplings to $c$ and $b$ quarks, such as diagrams with intermediate Higgs in the processes $c \bar c\to \gamma \gamma$ and $b \bar b \to \gamma \gamma$ and diagrams with the Higgs radiation from $c$ and $b$ quark lines in the associated production with $W$,$Z$ and $t \bar t$ (for example, $\bar s c \to W^+ \gamma \gamma$, $s \bar c \to W^- \gamma \gamma$, 6 diagrams) give very small yield and for this reason are omitted. Only gauge-invariant subset of 8 diagrams with Higgs boson radiation from the top line $gg \to \gamma \gamma t \bar t$ is calculated, omitting 18 diagrams with different topologies. Being marginally small in the SM, such amplitudes could give substantial contributions in the anomalous coupling space. We checked explicitly the absence of anomalous enhancements. 20 partonic subprocesses $q\# q\# \to q\# q\# \gamma \gamma$ were accounted for in the VBF channel, including interference terms between the diagrams. The notation $q\#$ is used to account for all possible combinations of $u$,$d$,$c$,$s$ quarks and anti-quarks.}
\item{{\it event signatures with four leptons} $gg\to \nu_l {\bar \nu}_l l^+ l^-$ and $gg\to l^+ l^- l^+ l^-$ including interference terms between $H\to W^+ W^-$ and $H\to ZZ$. Vector boson channels $H\to W^+ W^-$ and $H\to ZZ$ usually mix. A complete set of six Higgs production diagrams  in the channel $pp\to WW \to \nu_\mu {\bar \nu}_\mu \mu^+ \mu^-$ ($WW$ production via gluon fusion) is shown in Fig.\ref{ww}, where not only diagram 2 accounted for in the {\it production $\times$ decay} approximation contributes, but also $H\to ZZ$ channel, diagram 1, as well as four $s$-channel amplitudes, diagrams 4-6, together with interference terms which are rather small in this case. In this channel the $WW^*$-$ZZ^*$ interference term is negative (the value of the order of a few percent in comparison with $|W W^*|^2+|ZZ^*|^2$), cancelling the yield of $|ZZ^*|^2$ term. Leptonic event signatures in $WW$ and $ZZ$ VBF processes were included in the $(2\to 4$) $\times$ ($1\to 2$) infinitely small width approximation which can be justified by their smaller significance in comparison with $\gamma \gamma$ VBF.  One more example for the $ZZ\to \mu^+ \mu^- \mu^+ \mu^-$ channel is shown in Fig.\ref{zz}, where the "exchange" interference term $ZZ^*$-$ZZ^*$ is positive with the magnitude approximately 20\% of the $|ZZ^*|^2$ term. Diagrams with intermediate photons contribute insignificantly in the anomalous coupling space. As one could expect, the amplitudes with triple Higgs vertex (diagram 9) and $t$-channel gluon exchange (diagram 10) were found insignificant for the fit in the vicinity of the SM point. }
\item{{\it event signatures with $b \bar b$ and $\tau^+ \tau^-$}. For $H\to b \bar b$ the processes $q_1 \bar q_2 \to W b \bar b$ and $q \bar q \to Z b \bar b$ were calculated. Again  diagrams with Higgs boson radiation from $c$ and $b$ quark lines were neglected. For $H\to \tau^+ \tau^-$ channel we calculated $\tau^+ \tau^-$, $\tau^+ \tau^-$ VBF, $\tau^+ \tau^- t \bar t$, $\tau^+ \tau^- W$ and $\tau^+ \tau^- Z$ production.}
\end{itemize}

\begin{figure}[h]
\begin{minipage}[c]{.50\textwidth}
\includegraphics[width=3in,height=2in]{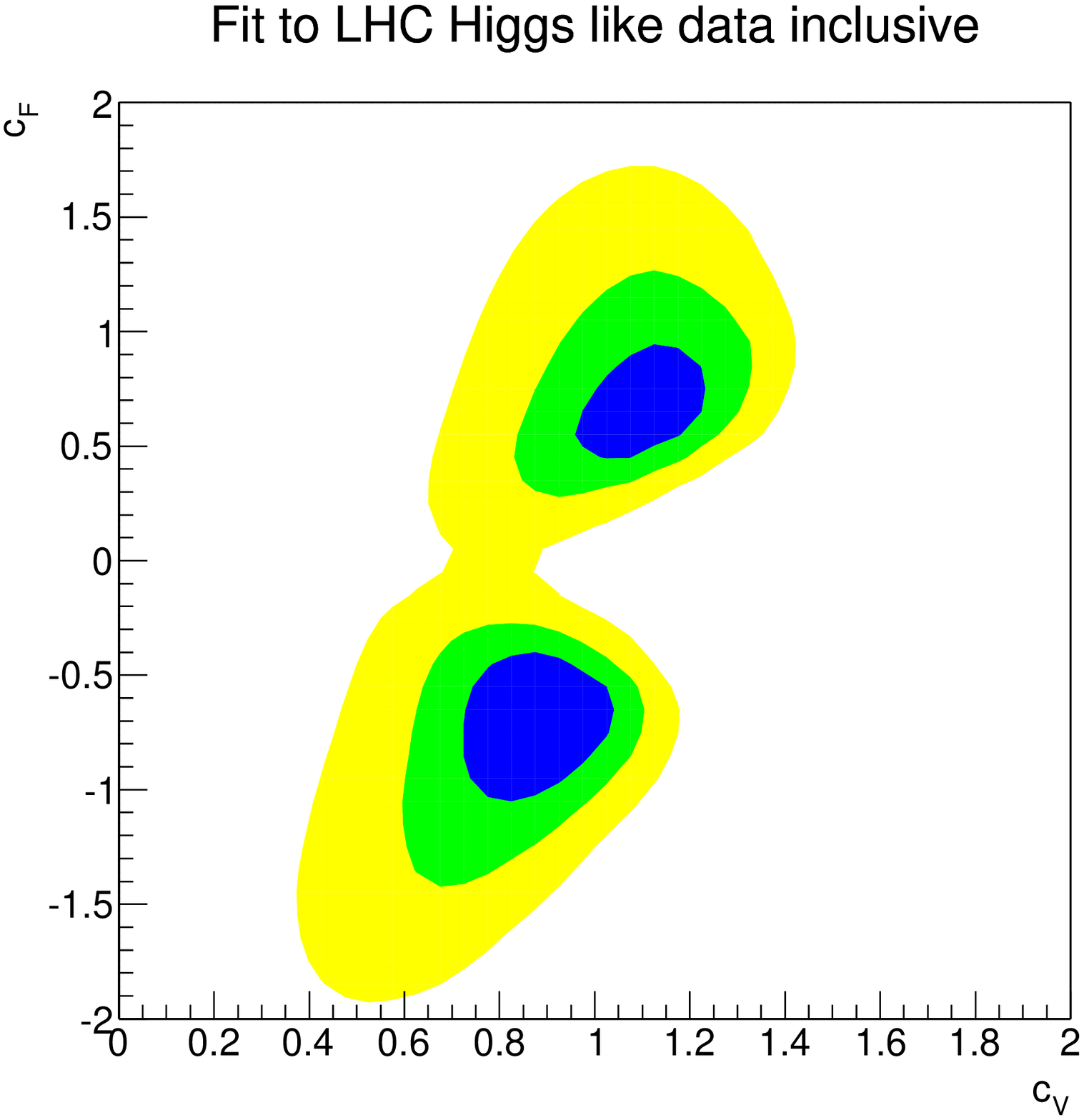}
\begin{center}
{\it (a)}
\end{center}
\end{minipage}
\begin{minipage}[c]{.50\textwidth}

\includegraphics[width=3in]{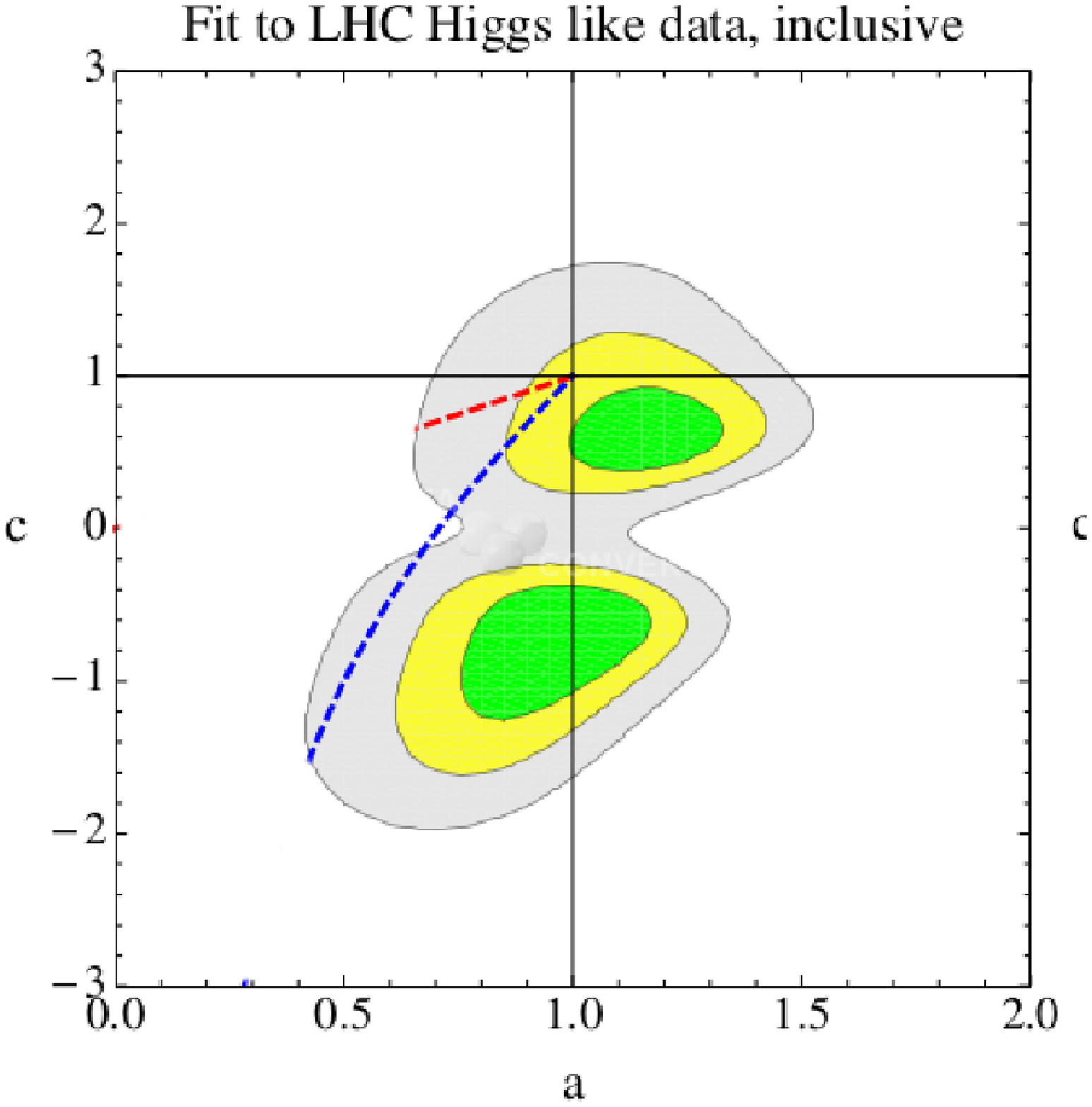}
\begin{center}
{\it (b)}
\end{center}
\end{minipage}

\vspace{10mm}
\caption[]{\label{our_espinosa} \it (a) - global $\chi^2$ fit in the $(c_V,c_F)$ plane calculated with Higgs boson width for all two-particle, $WW^*$ and $ZZ^*$ decay channels including VBF(diagrams with gluon fusion omitted) combined with $\gamma \gamma$ VBF, within the production$\times$decay approximation, (b) - global $\chi^2$ fit in the $(a,c)$ plane from \cite{fit1} }

\end{figure}

\begin{figure}[h]
\begin{minipage}[c]{.30\textwidth}

\includegraphics[width=2.0in]{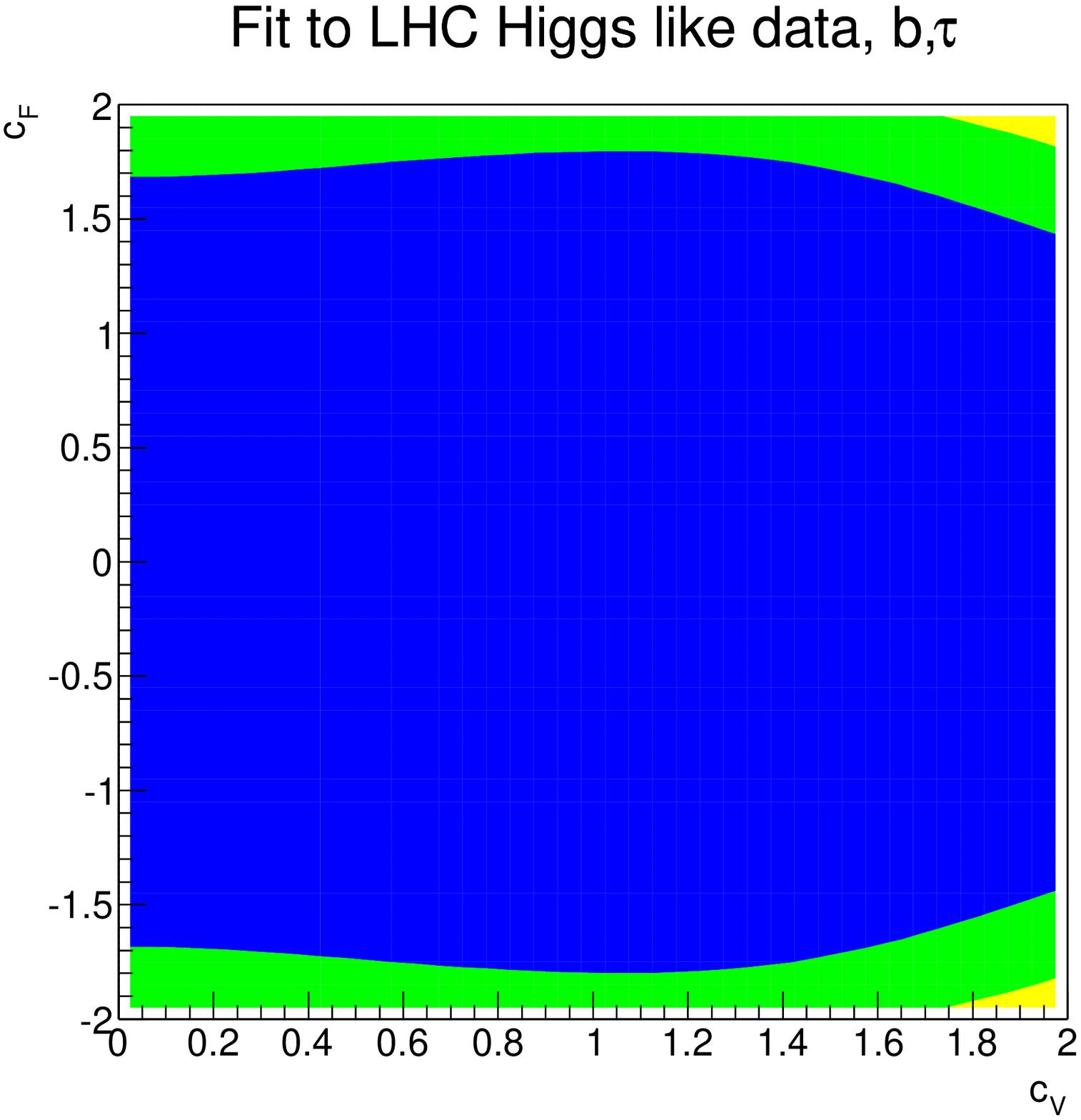}
\begin{center}
{\it (a)}
\end{center}
\end{minipage}
\hspace{0.5cm}\begin{minipage}[c]{.30\textwidth}

\includegraphics[width=2.0in]{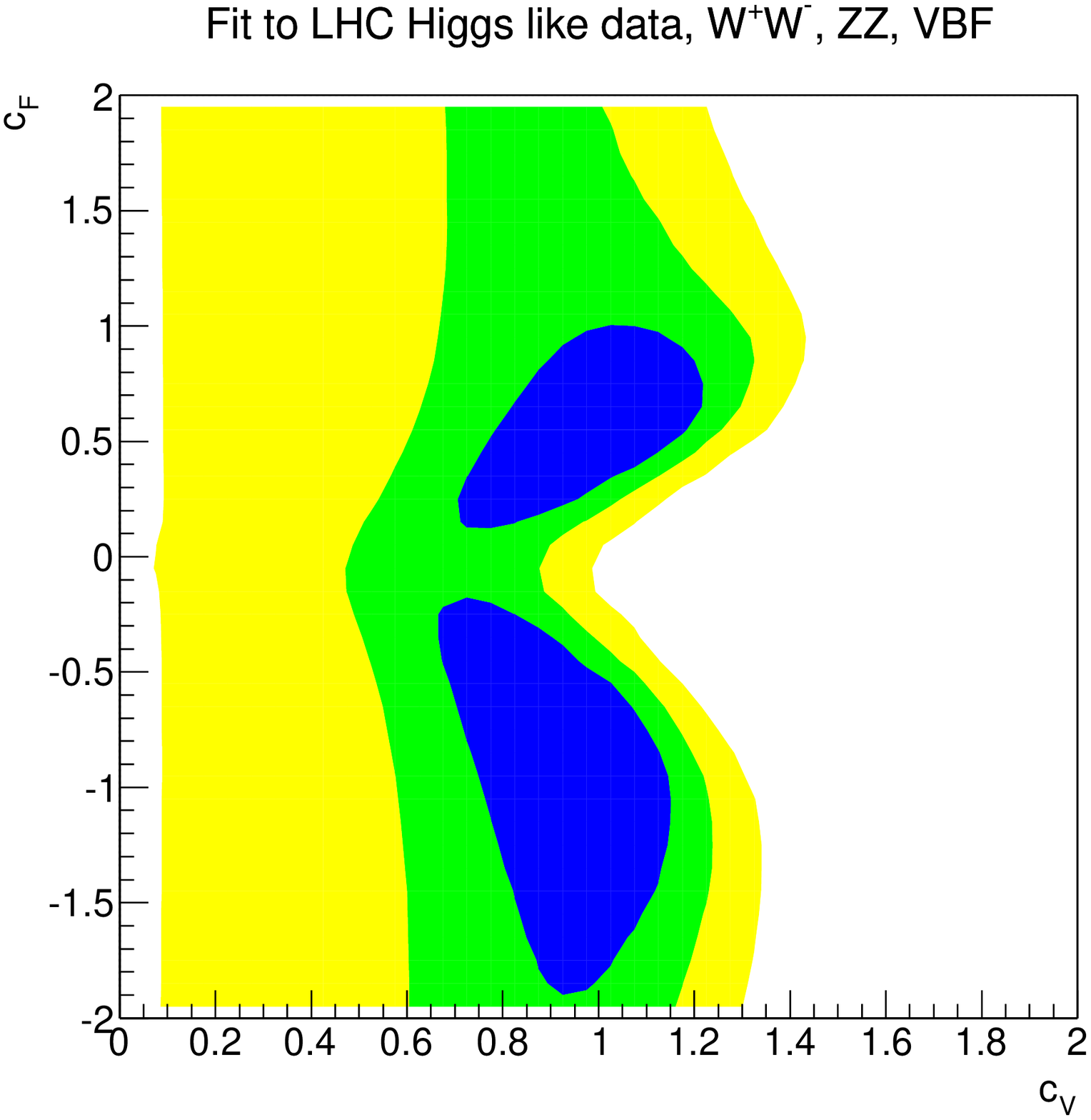}
\begin{center}
{\it (b)}
\end{center}
\end{minipage}
\hspace{0.5cm} \begin{minipage}[c]{.30\textwidth}

\includegraphics[width=2.0in]{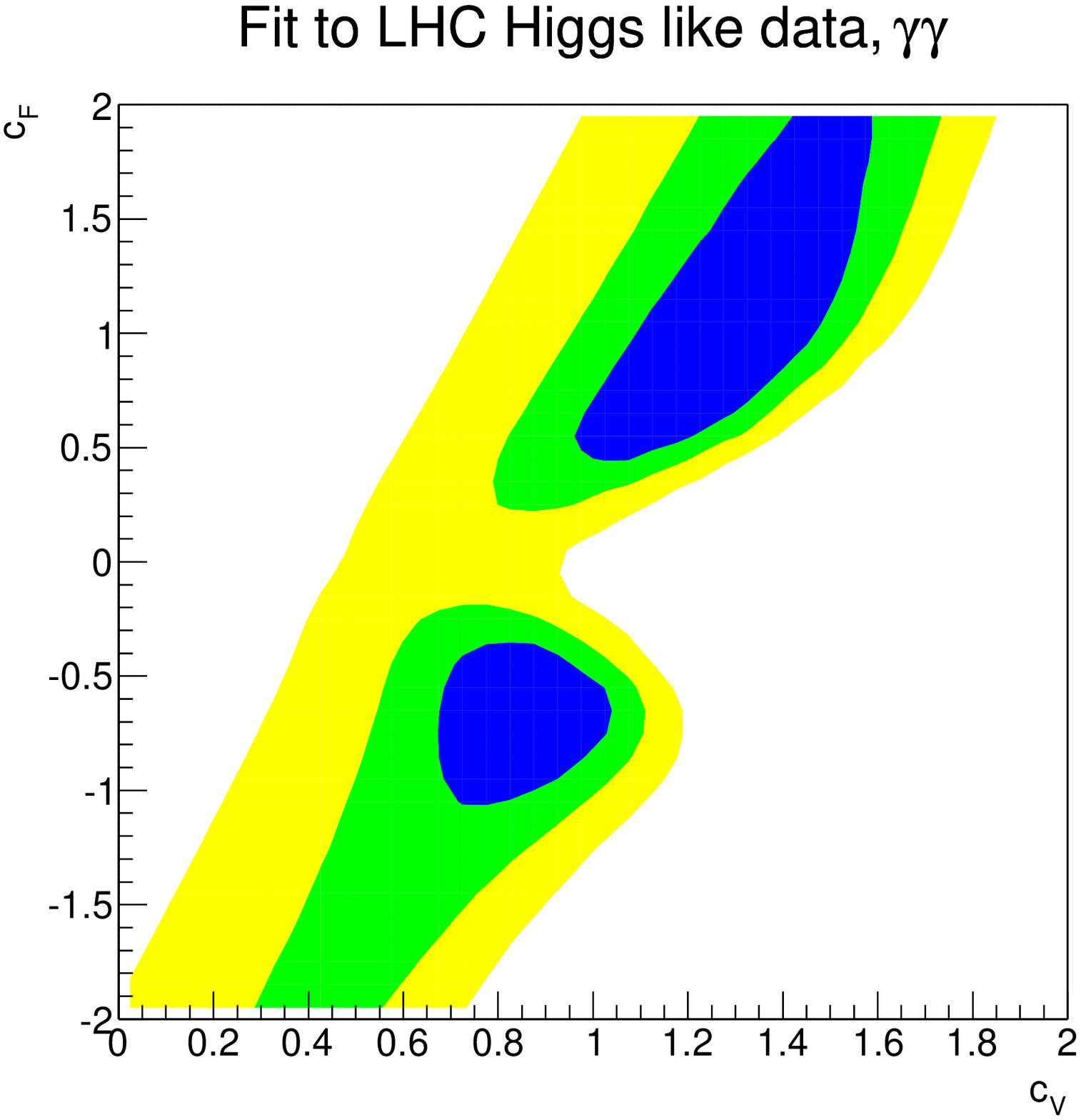}
\begin{center}
{\it (c)}
\end{center}
\end{minipage}

\begin{minipage}[c]{.7\textwidth}

\includegraphics[width=6.0in]{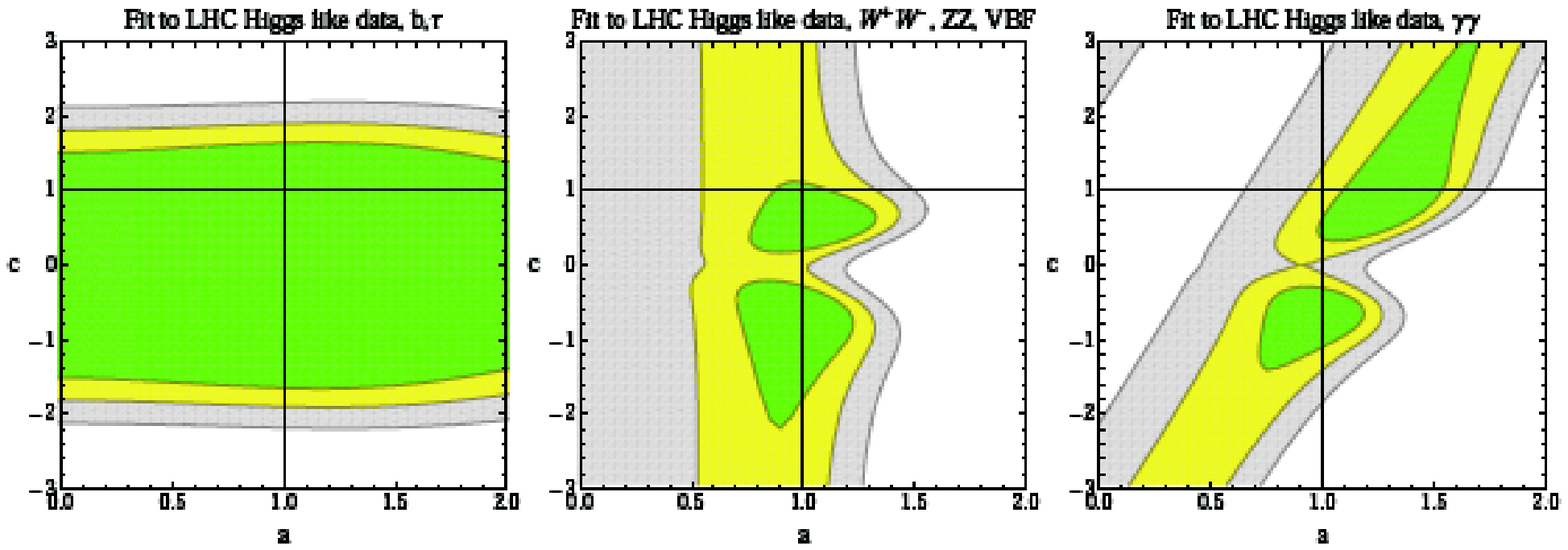}
\vspace{-0.8cm}
\begin{center}
{\it \begin{tiny}
•
\end{tiny} \hskip 3.5cm (d)}
\end{center}

\end{minipage}

\vspace{10mm}
\caption[]{\label{isw_threechannels} \it $\chi^2$ fits in the $(c_V,c_F)$ plane calculated within the 
production $\times$ decay approximation (a) - $b \bar b$ and $\tau^+ \tau^-$ channels; (b) - $WW^*$ and $ZZ^*$ channels including VBF (diagrams with gluon fusion omitted) combined with $\gamma \gamma$ VBF, (c) - $\gamma \gamma$ channels including VBF; (d) -  the same fits for $b \bar b$, $\tau^+ \tau^-$, $WW^*$, $ZZ^*$ and $\gamma \gamma$ channels in the $(a,c)$ plane from \cite{fit1}. Note that different ranges for $c_F$ are used in upper and lower rows of plots. }

\end{figure}

For validation of our codes and numerical fitting procedures we reproduce the global fit of the first paper in \cite{fit1}, which was reconstructed in the ($a$,$c$) plane on the basis of 2012 post-Moriond data in the infinitely small width approximation (ISW approximation); compare Fig.\ref{our_espinosa}(a) and Fig.\ref{our_espinosa}(b). Our contours in the ($c_V$,$c_F$) plane were generated also in the ISW approximation.  For example, within the ISW for the process $pp \to \nu_\mu {\bar \nu}_\mu \mu^+ \mu^-$ first the $2\to 3$ process $pp\to Z \mu^+ \mu^-$ is calculated on a grid 31$\times$31 in the ($c_W$,$c_F$) plane, which is then convoluted with the branching $Z\to \mu^+ \mu^-$ calculated on the same grid\footnote{A special regime of 'table calculations' (numerical operations with multidimensional tables) has been implemented in CompHEP version 4.5.}. Fig.\ref{isw_threechannels} demonstrates good agreement in the ISW approximation for the three groups of channels with: (a) $H\to \gamma \gamma$ in the final state, (b) $H\to WW^*$ and $H\to ZZ^*$, (c) $H\to b {\bar b}$ and $H\to \tau^+ \tau^-$. 

\begin{figure}[h]
\begin{minipage}[c]{.30\textwidth}

\includegraphics[width=2.0in]{fig4a.eps}
\begin{center}
{\it (a)}
\end{center}
\vspace*{0.0cm}
\end{minipage}
\hspace{0.5cm}\begin{minipage}[c]{.30\textwidth}

\includegraphics[width=2.0in]{fig4b.eps}
\begin{center}
{\it (b)}
\end{center}
\vspace*{-0.0cm}
\end{minipage}
\hspace{0.5cm} \begin{minipage}[c]{.30\textwidth}

\includegraphics[width=2.0in]{fig4c.eps}
\begin{center}
{\it (c)}
\end{center}
\vspace*{-0.0cm}
\end{minipage}

\begin{minipage}[c]{.30\textwidth}

\includegraphics[width=2.0in]{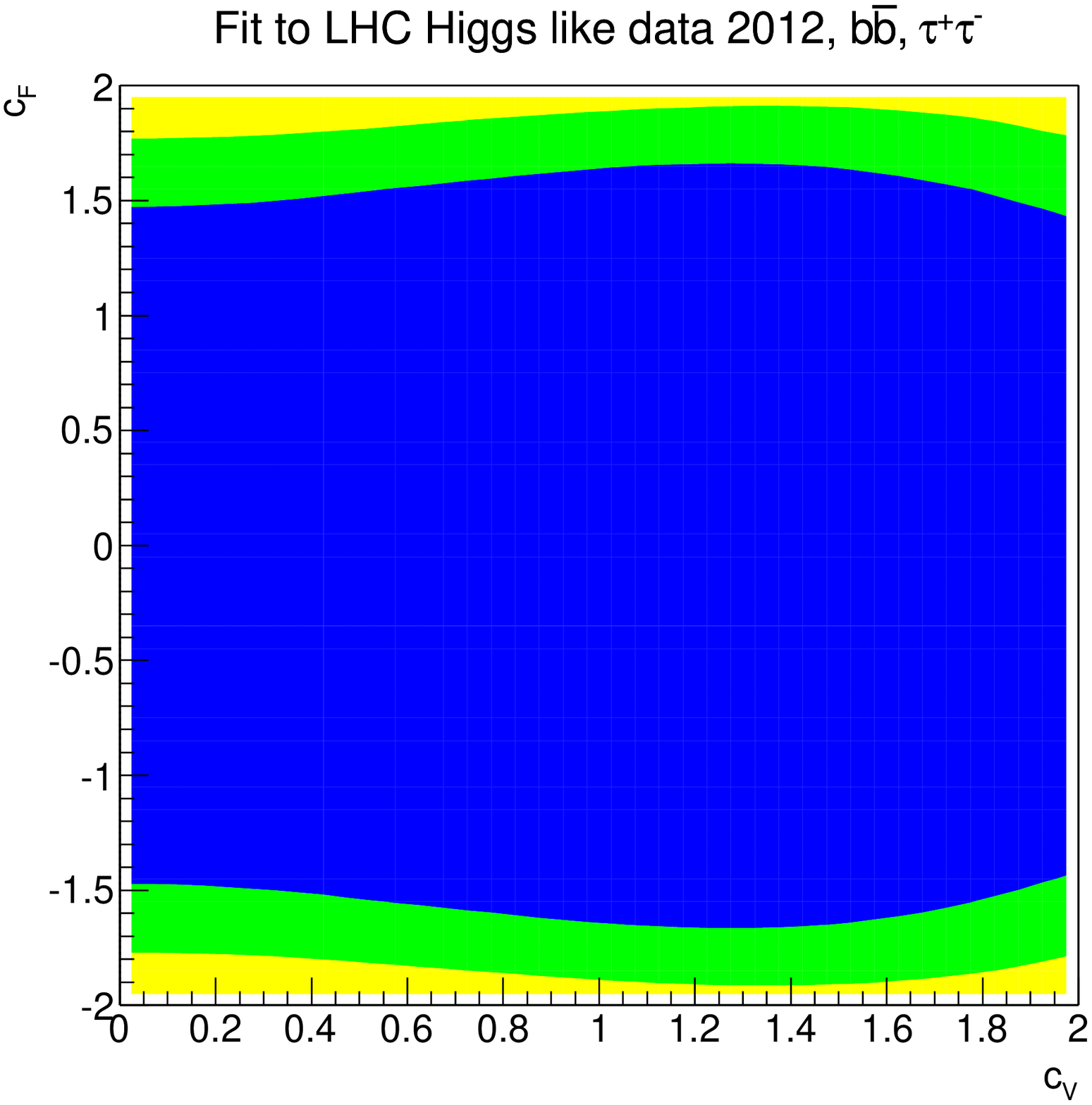}
\begin{center}
{\it (d)}
\end{center}
\vspace*{0.0cm}
\end{minipage}
\hspace{0.5cm}\begin{minipage}[c]{.30\textwidth}

\includegraphics[width=2.0in]{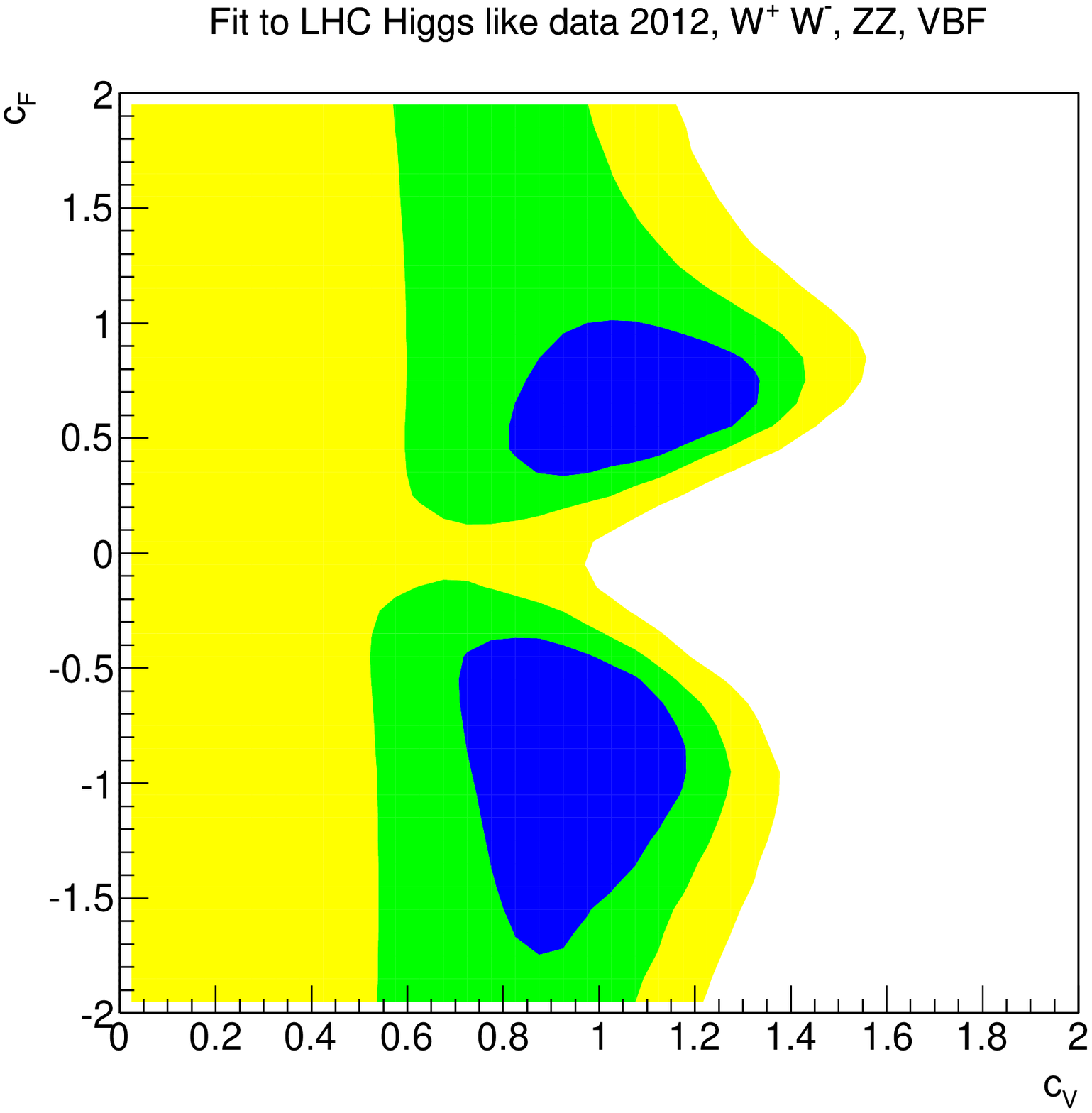}
\begin{center}
{\it (e)}
\end{center}
\vspace*{-0.0cm}
\end{minipage}
\hspace{0.5cm} \begin{minipage}[c]{.30\textwidth}

\includegraphics[width=2.0in]{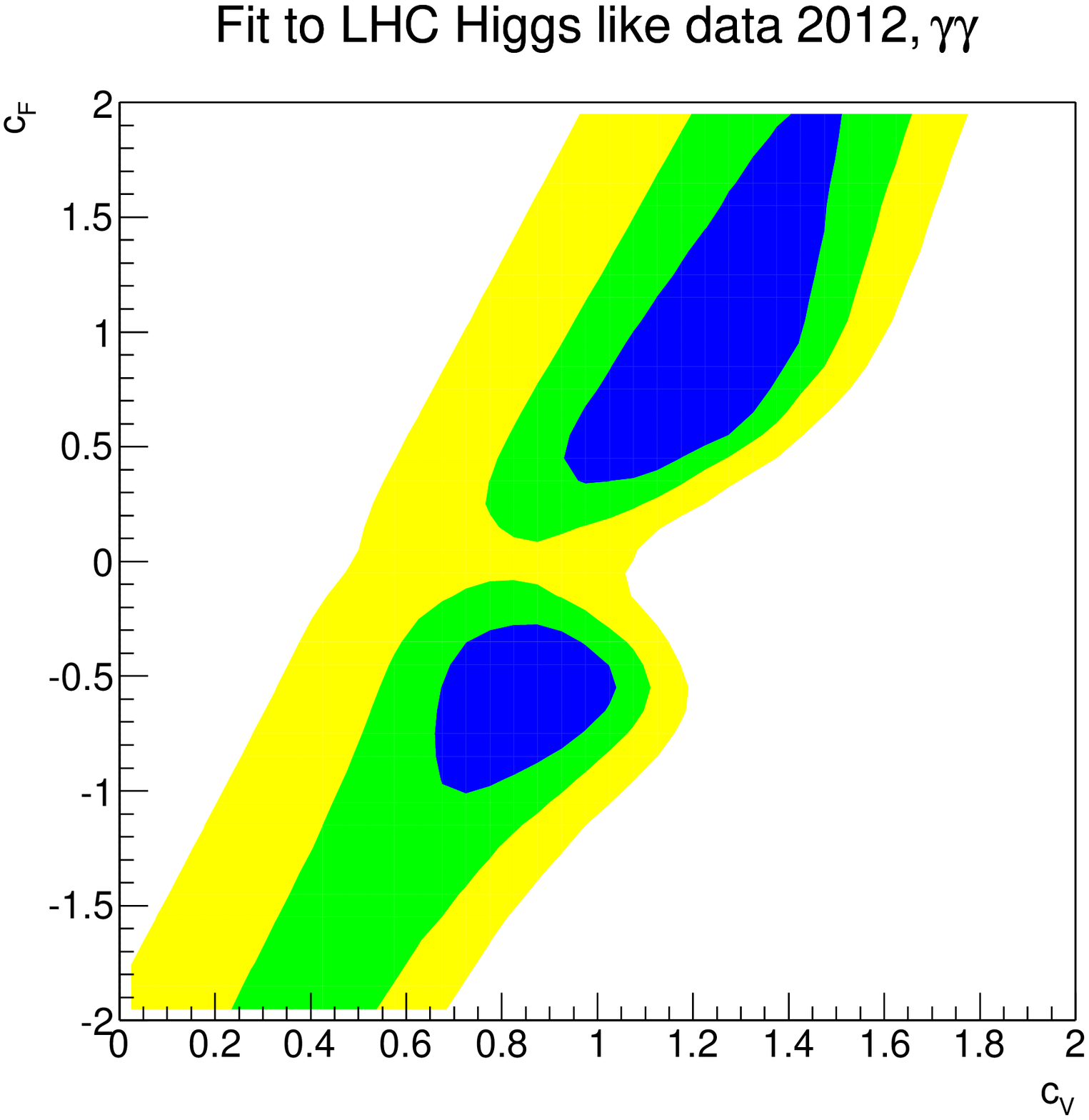}
\begin{center}
{\it (f)}
\end{center}
\vspace*{-0.0cm}
\end{minipage}

\vspace{10mm}
\caption[]{\label{ctl_threechannels} (a),(b),(c) - \it $\chi^2$ fits (2012 data) in the $(c_V,c_F)$ plane calculated within the 
production $\times$ decay approximation; (d),(e),(f) - the same fits calculated with complete gauge invariant sets of diagrams. (a),(d) - $b \bar b$ and $\tau^+ \tau^-$ channels; (b),(e) - $WW^*$ and $ZZ^*$ channels including VBF (diagrams with gluon fusion omitted) combined with $\gamma \gamma$ VBF, (c),(f) - $\gamma \gamma$ channels including VBF.}

\end{figure}

\begin{figure}[h]
\begin{minipage}[c]{.50\textwidth}

\includegraphics[width=2.5in]{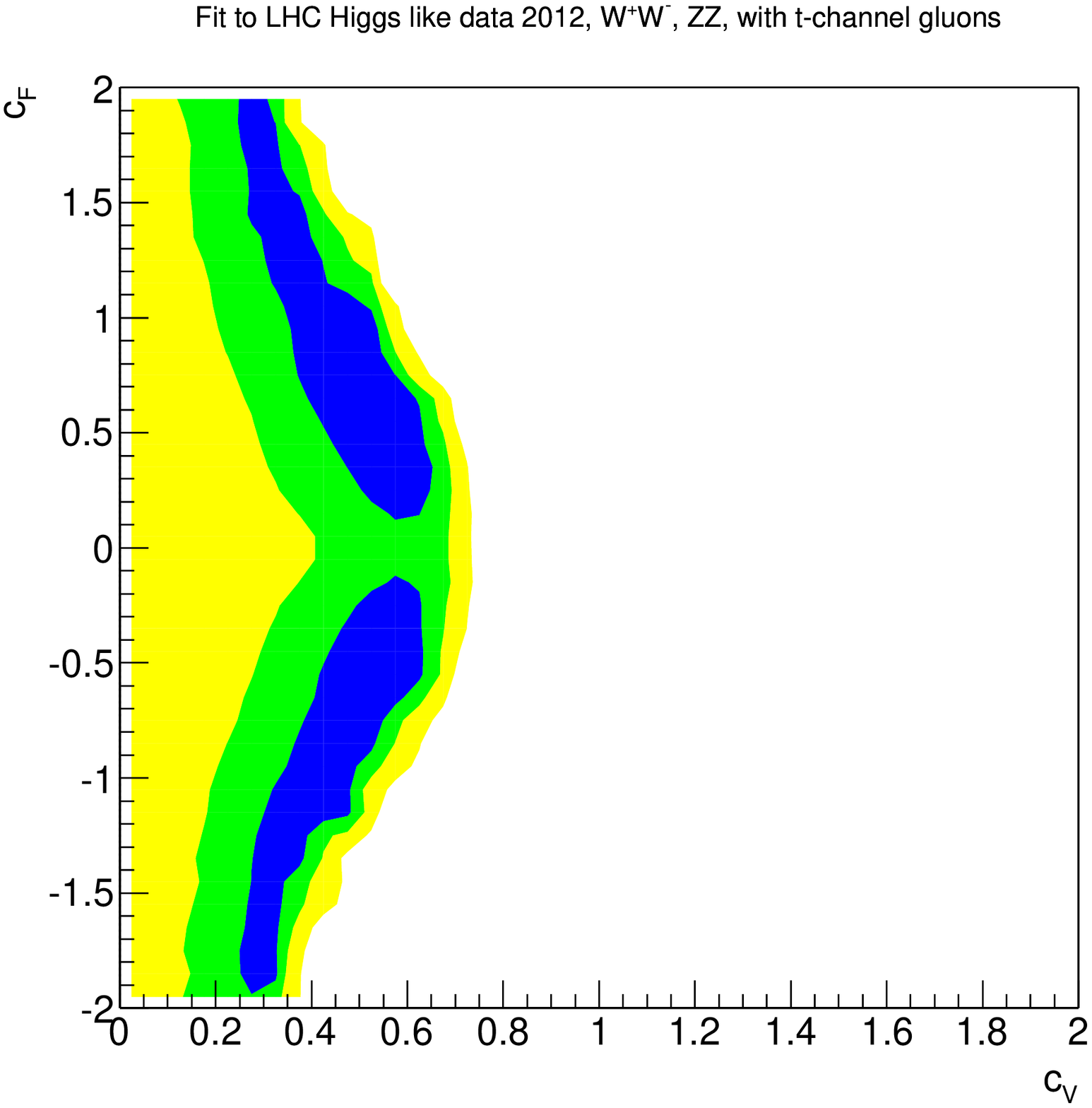}
\vspace{-0.3cm}
\begin{center}
{\it (a)}
\end{center}
\vspace*{0.0cm}
\end{minipage}
\hspace{0.5cm}\begin{minipage}[c]{.50\textwidth}

\includegraphics[width=2.5in]{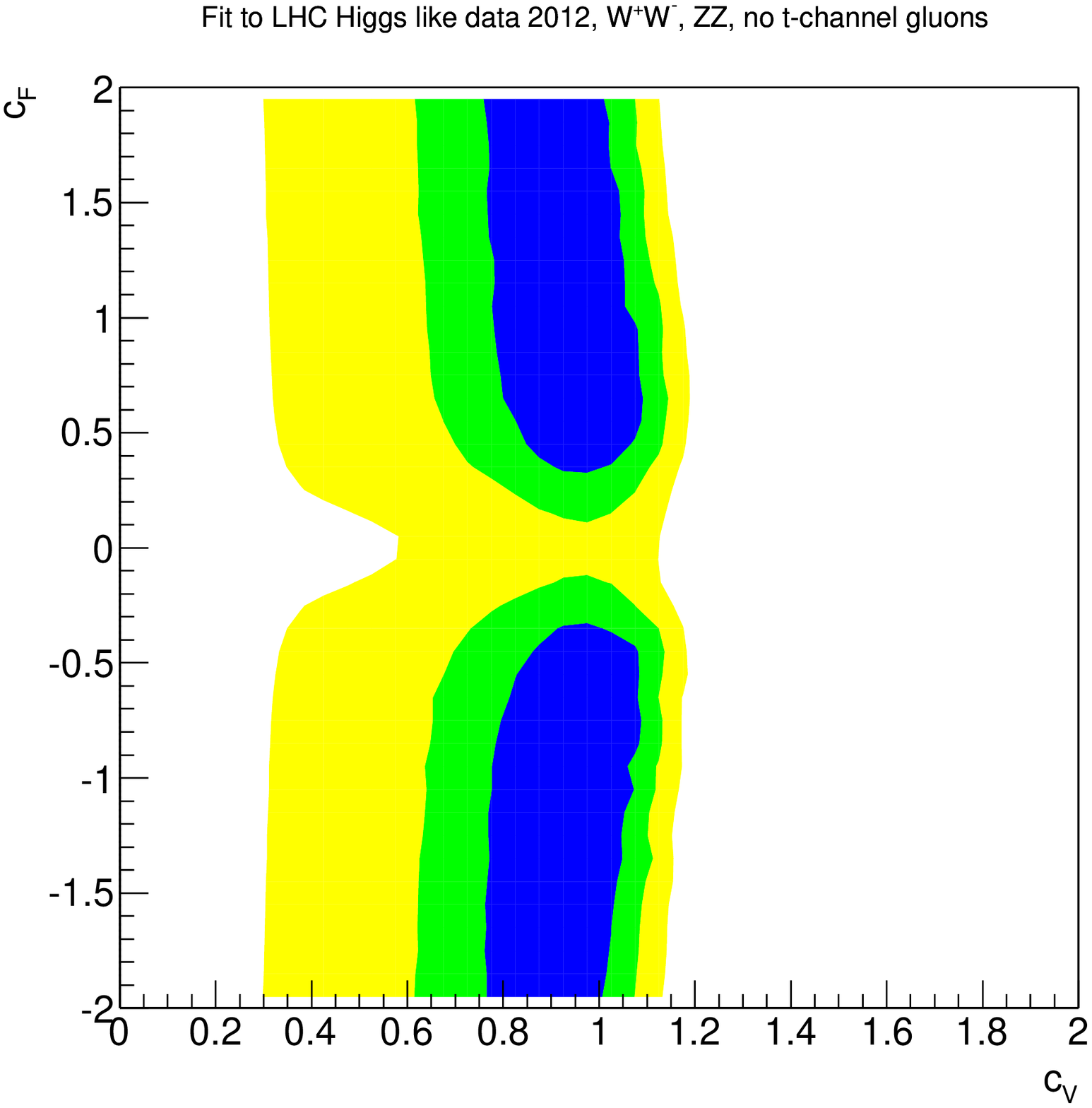}
\vspace{-0.3cm}
\begin{center}
{\it (b)}
\end{center}
\vspace*{-0.0cm}
\end{minipage}

\vspace{10mm}
\caption[]{\label{with_no_vbf_glue} \it $\chi^2$ fits in the $(c_V,c_F)$ plane calculated for $WW^*$ and $ZZ^*$ channels, \hskip 3mm (a) - including VBF (ladder) diagrams with the fusion of gluons radiated from the quark lines, (b) - diagrams with intermediate gluons omitted. Identical signal strength and signal strength error were taken for the four-lepton final states produced either without forward jets or with forward jets tagging. $\gamma \gamma$ VBF diagrams are not accounted for.
}

\end{figure}

According to Fig.\ref{our_espinosa} the data is consistent with the standard Higgs sector hypothesis at the 82\% CL. The symmetry $c_F \to -c_F$ for for preferrable regions in the ($c_V$,$c_F$) plane is violated not only by loop corrections but also beyond the infinitely small width approximation, where a number of additional diagrams and interference terms appear. For example, when the abovementioned $pp \to \nu_\mu {\bar \nu}_\mu \mu^+ \mu^-$ is calculated beyond the ISW as $2\to 4$ process at complete tree level, both $H\to ZZ\to \nu_\mu {\bar \nu}_\mu \mu^+ \mu^- $ and $H\to W^+ W^- \to \nu_\mu {\bar \nu}_\mu \mu^+ \mu^-$ with interference between them (see Fig.\ref{ww}) contribute. Fig.\ref{ctl_threechannels} demonstrates visible deviations of the exclusion contours for the abovementioned group (b). Important additional source of deviation could be diagrams with gluon fusion (gluons radiated from the quark lines). The role of such diagrams is illustrated separately in Fig.\ref{with_no_vbf_glue}. Gluon fusion, accounted for in the $Q^2$ evolution of the parton distribution functions for the fully inclusive processes, deserves however careful evaluation if some specific selection criteria for the forward jets is used. In evaluations we omit gluon fusion amplitudes.  

\begin{figure}[h]

. \hspace{-1.5cm}
\begin{minipage}[c]{.45\linewidth}

\includegraphics[width=3.2in]{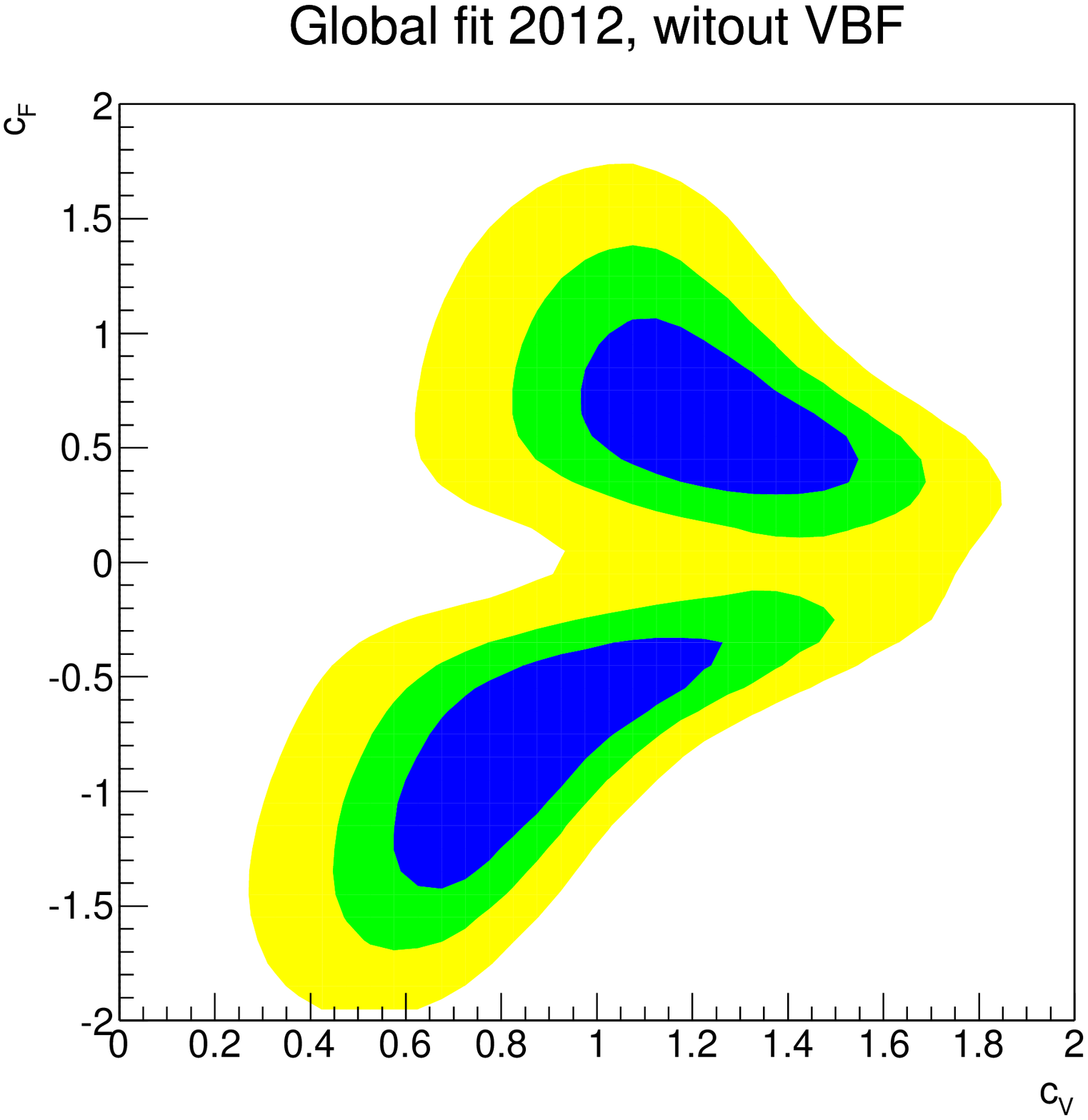}
\vspace{-1.0cm}
\begin{center}
{\it $ $ \vspace{4mm} \hspace{8mm}(a)}
\end{center}
\end{minipage}
\hspace{8mm} \begin{minipage}[c]{.45\linewidth}

\includegraphics[width=3.2in]{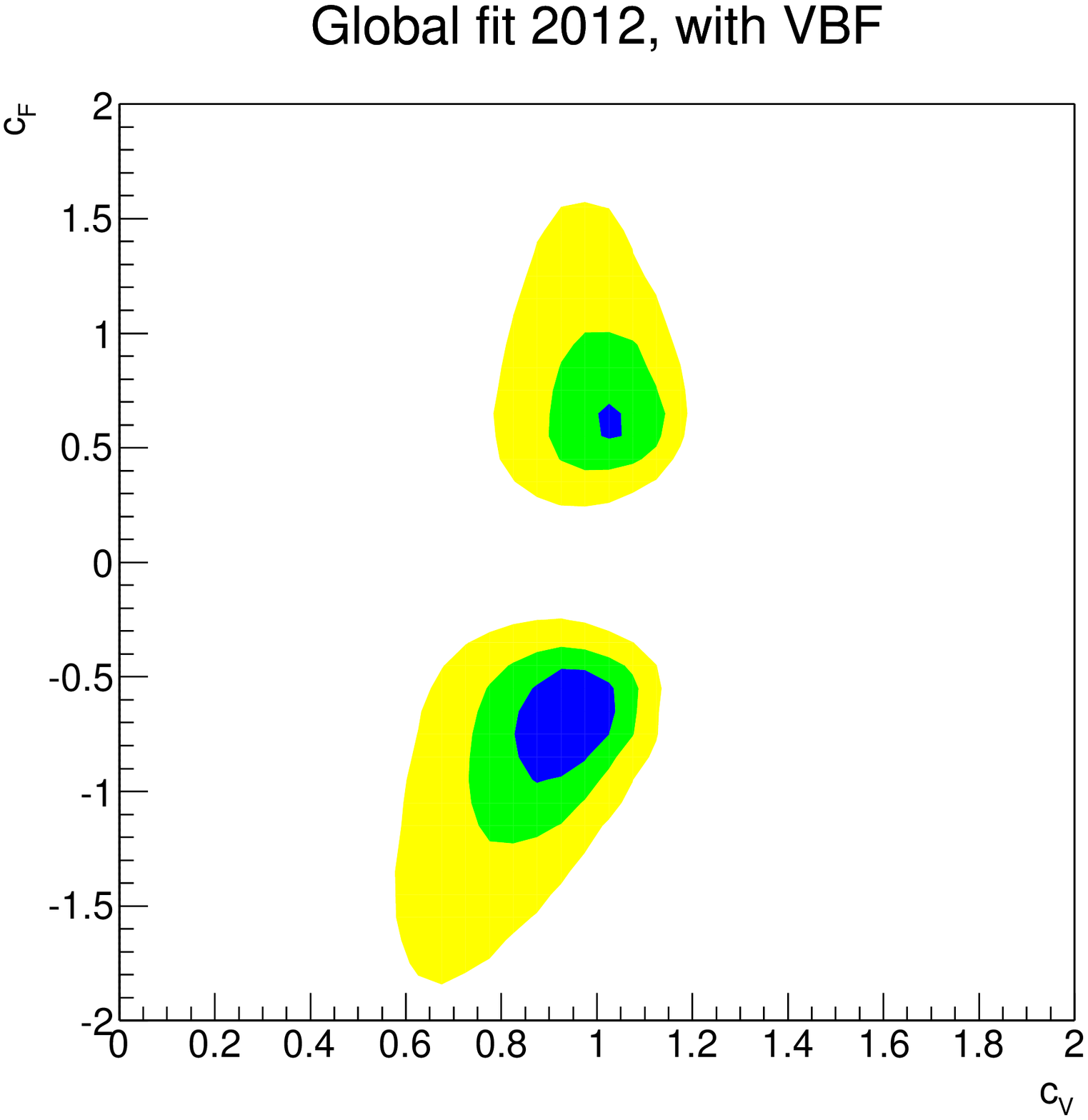}
\vspace{-1.0cm}
\begin{center}
{\it $ $ \vspace{4mm} \hspace{12mm} (b)}
\end{center}
\end{minipage}

. \hspace{-1.5cm}
\begin{minipage}[c]{.45\linewidth}

\includegraphics[width=3.2in]{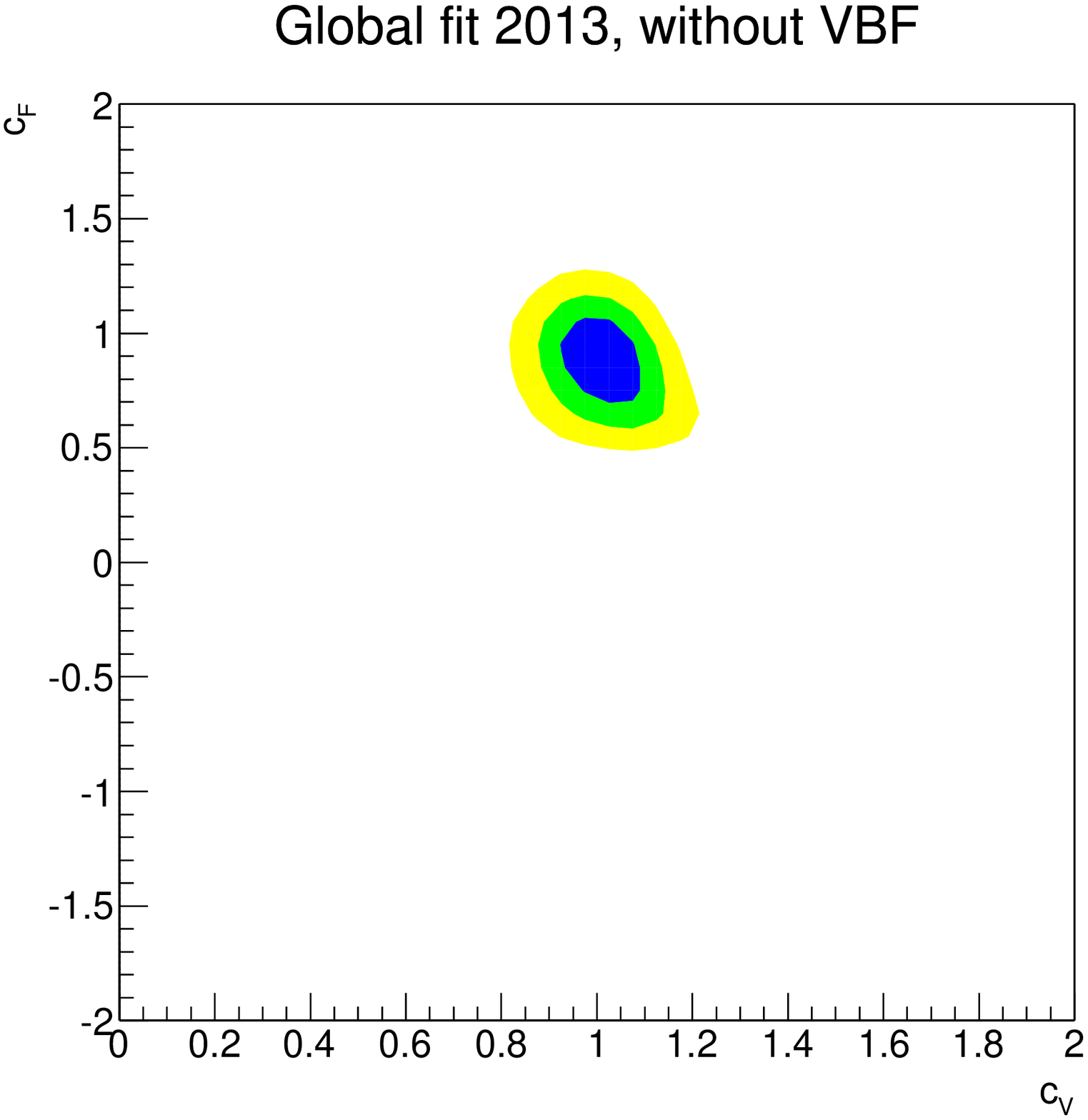}
\vspace{-1.2cm}
\begin{center}
{\it $ $ \hspace{8mm}(c)}
\end{center}
\end{minipage}
\hspace{8mm} \begin{minipage}[c]{.45\linewidth}

\includegraphics[width=3.2in]{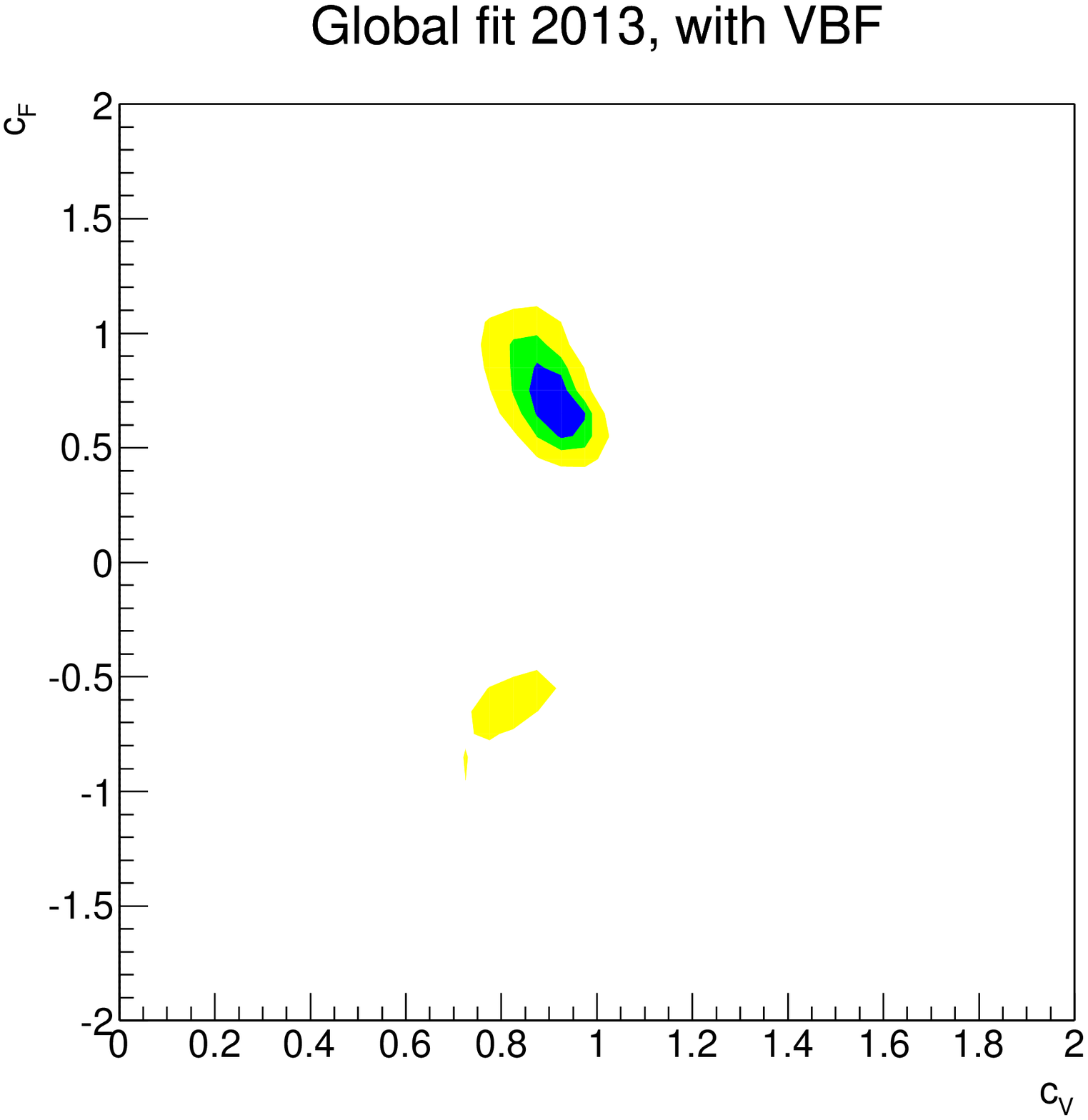}
\vspace{-1.2cm}
\begin{center}
{\it $ $ \hspace{12mm} (d)}
\end{center}
\end{minipage}

\vspace{10mm}
\caption[]{\label{global} Global \it $\chi^2$ fits in the $(c_V,c_F)$ plane. (a),(c) - calculated without VBF diagrams in the $\gamma \gamma$, $WW^*$ and $ZZ^*$ channels, (b),(d) - calculated with VBF diagrams in the $\gamma \gamma$, $WW^*$ and $ZZ^*$ channels; (a),(b) based on 2012 data, (c),(d) based on preliminary 2013 data (Table \ref{lc2013}).  }

\end{figure}

The relevance of VBF in the $H\to \gamma \gamma$ channel is demonstrated in Fig.\ref{global} (upper row of plots). In the three-dimensional plot $\sigma$($c_W$,$c_F$) the surfaces of VBF cross section and the cross section of processes without forward jets have opposite behavior, while one is increasing the other is decreasing, with the result for the exclusion contour very sensitive to the precision of evaluations.    

\begin{table}
\begin{center}
\begin{tabular}{|l|l|l|} \hline
channel & ATLAS & \hskip 2mm CMS \\ \hline
$VH\to Vb \bar b$ & -0.4$\pm$1.0 & 1.15$\pm$0.62 \\  
$H\to \tau^+ \tau^-$ & 0.8$\pm$0.7 & 1.10$\pm$0.41 \\
$H\to W W^*$ & 1.0$\pm$0.3 & 0.68$\pm$0.20\\
$H\to Z Z^*$ & 1.5$\pm$0.4 & 0.92$\pm$0.28\\
$H\to \gamma \gamma$ & 1.6$\pm$0.3 & 0.77$\pm$0.27 \\ \hline
\end{tabular}
\end{center}
\caption{\label{lc2013} The signal strength and the signal strength error following \cite{lc2013}.}
\end{table}

\begin{figure}[h]
\begin{minipage}[c]{.30\textwidth}

\includegraphics[width=2.1in]{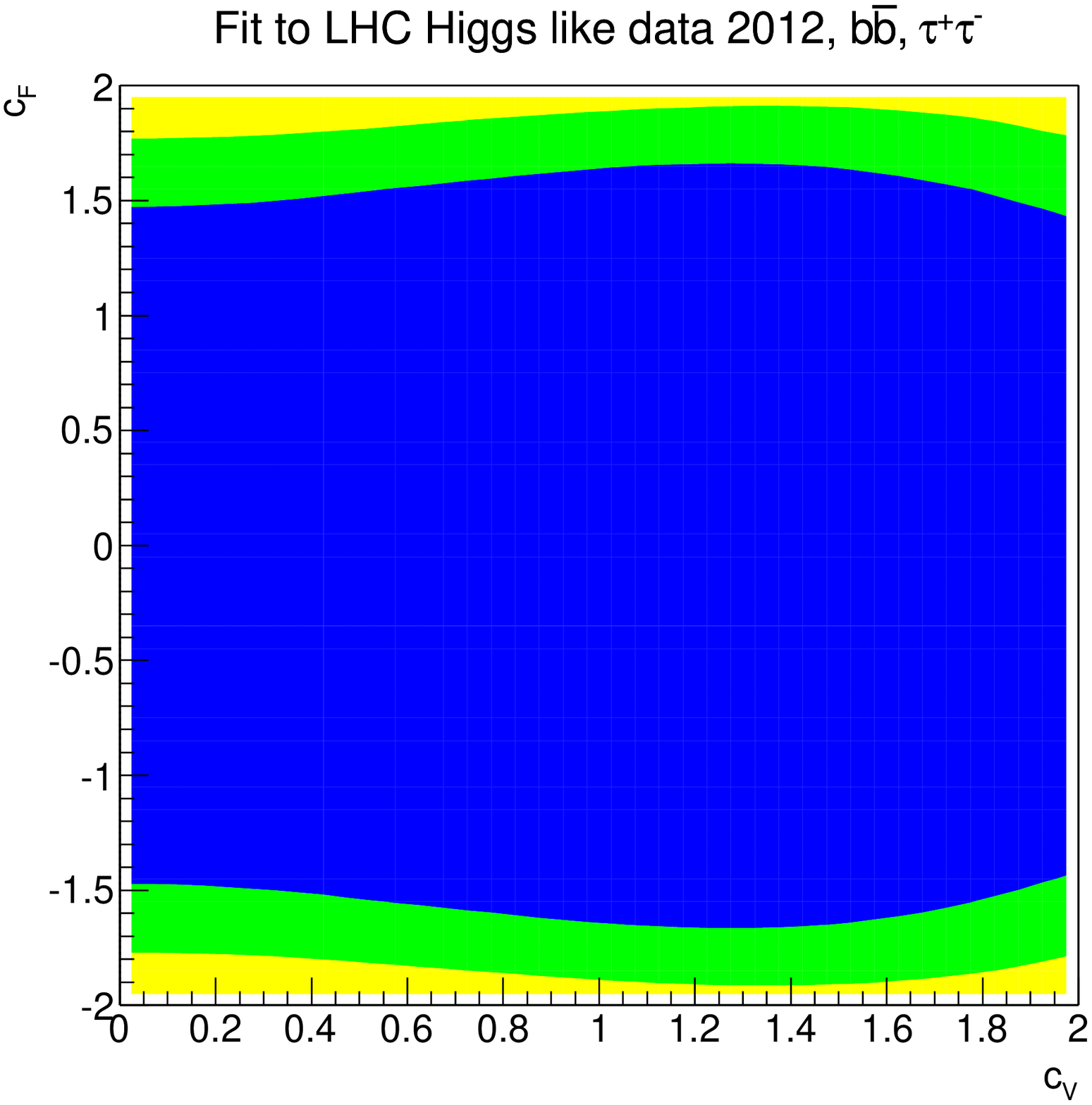}
\begin{center}
{\it (a)}
\end{center}
\vspace*{0.0cm}
\end{minipage}
\hspace{0.5cm}\begin{minipage}[c]{.30\textwidth}

\includegraphics[width=2.1in]{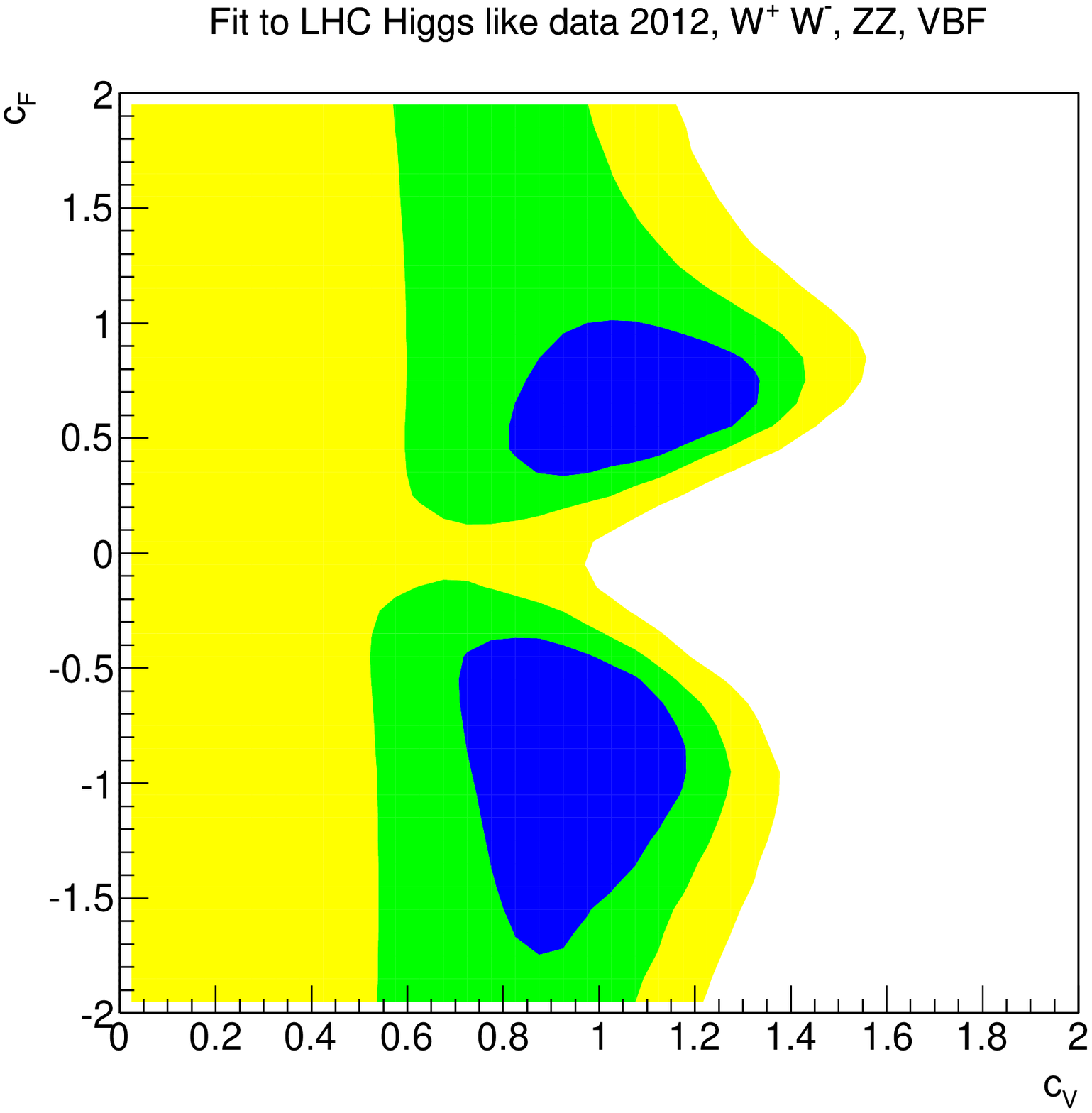}
\begin{center}
{\it (b)}
\end{center}
\vspace*{-0.0cm}
\end{minipage}
\hspace{0.5cm} \begin{minipage}[c]{.29\textwidth}

\includegraphics[width=2.1in]{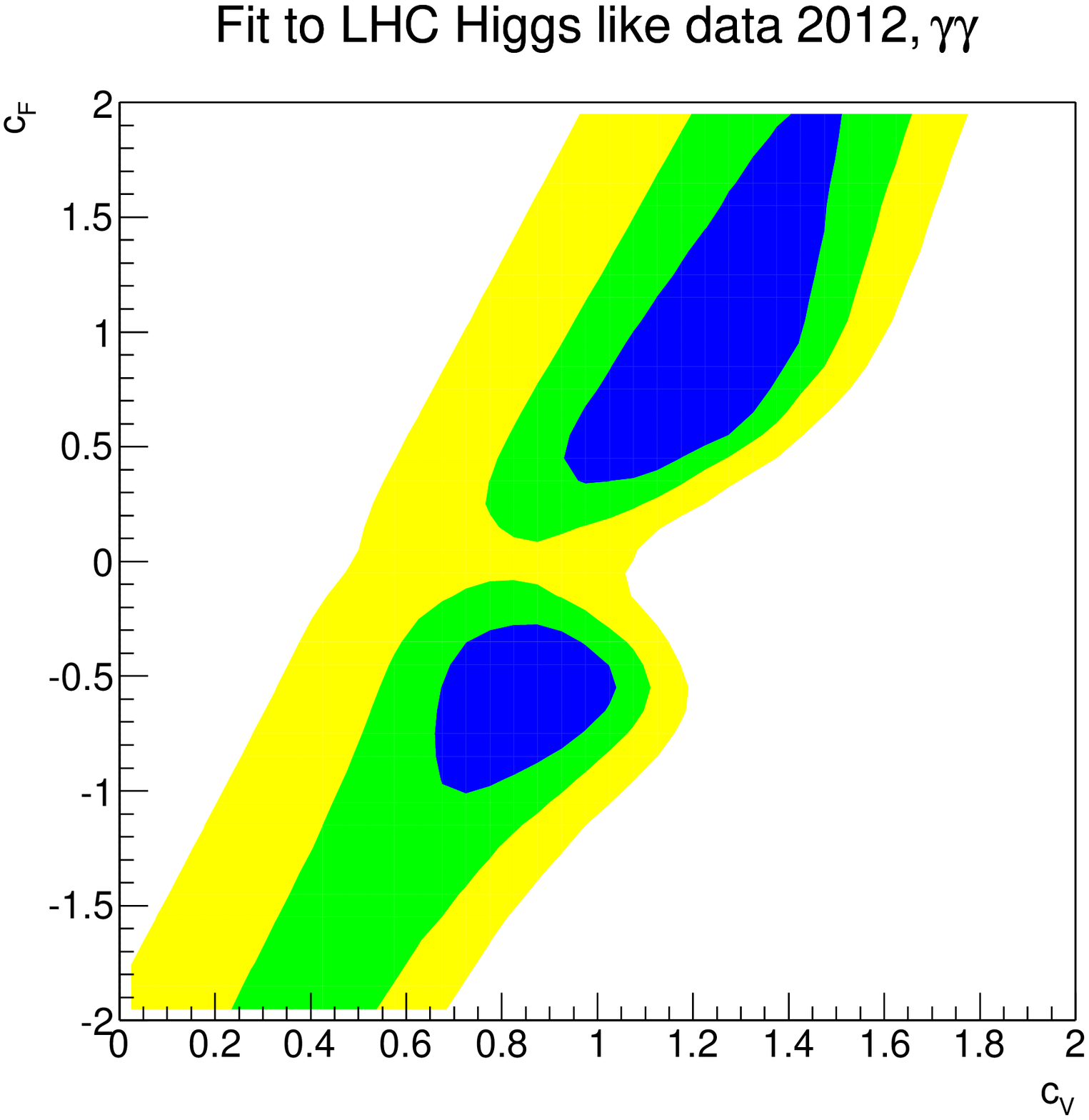}
\begin{center}
{\it (c)}
\end{center}
\vspace*{-0.0cm}
\end{minipage}

\begin{minipage}[c]{.30\textwidth}

\includegraphics[width=2.1in]{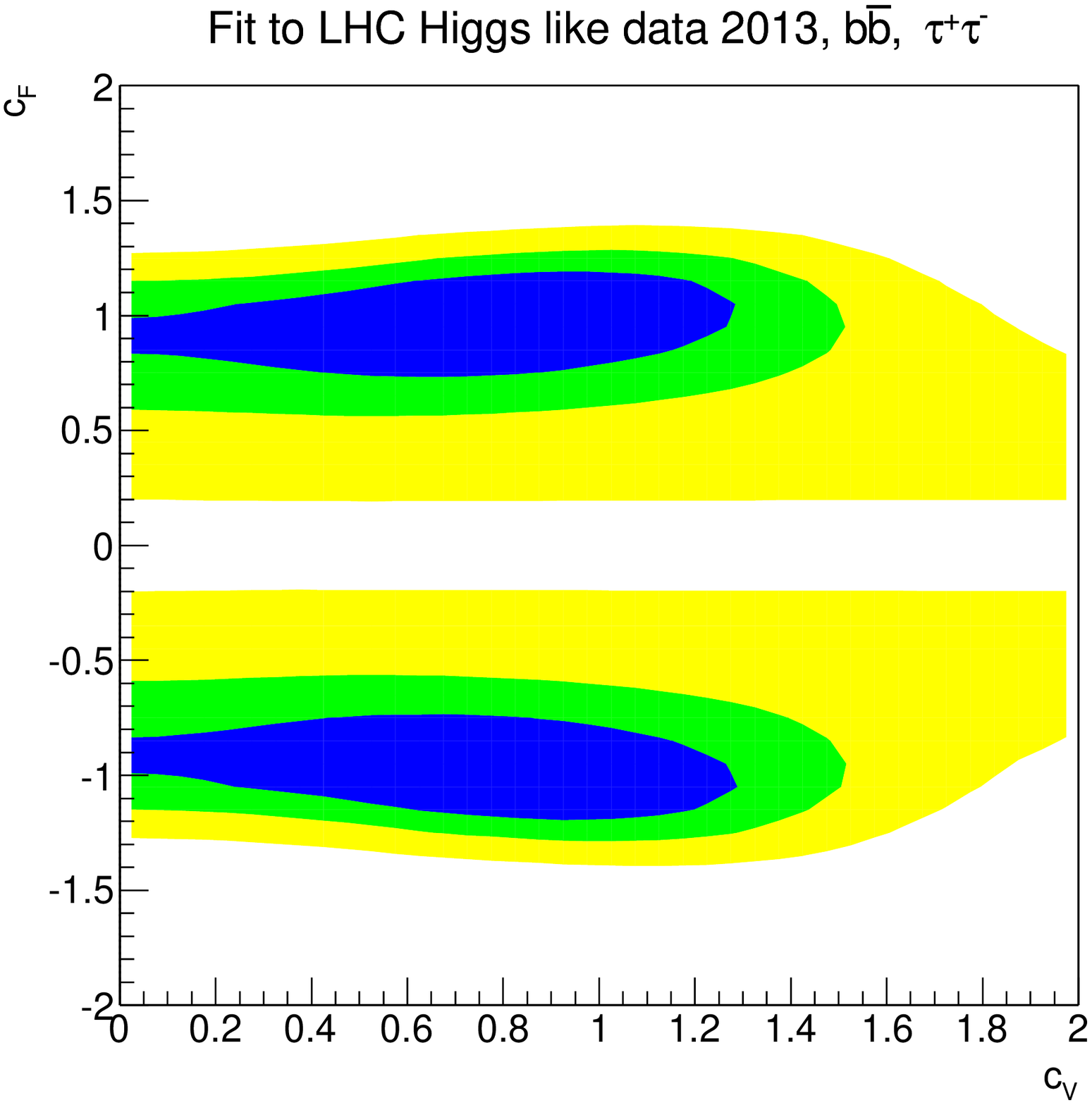}
\begin{center}
{\it (d)}
\end{center}
\vspace*{0.0cm}
\end{minipage}
\hspace{0.5cm}\begin{minipage}[c]{.30\textwidth}

\includegraphics[width=2.1in]{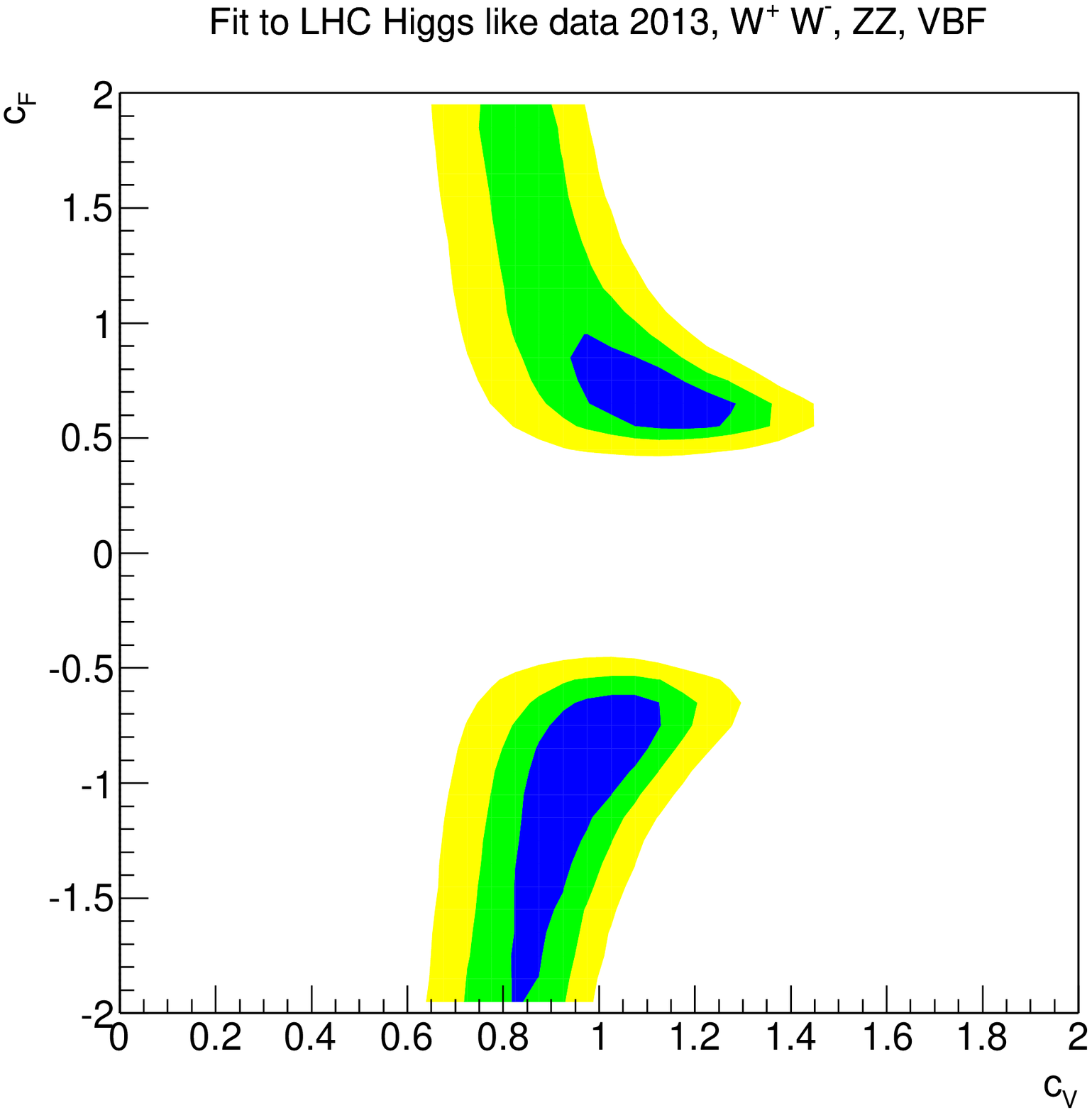}
\begin{center}
{\it (e)}
\end{center}
\vspace*{-0.0cm}
\end{minipage}
\hspace{0.5cm} \begin{minipage}[c]{.30\textwidth}

\includegraphics[width=2.05in]{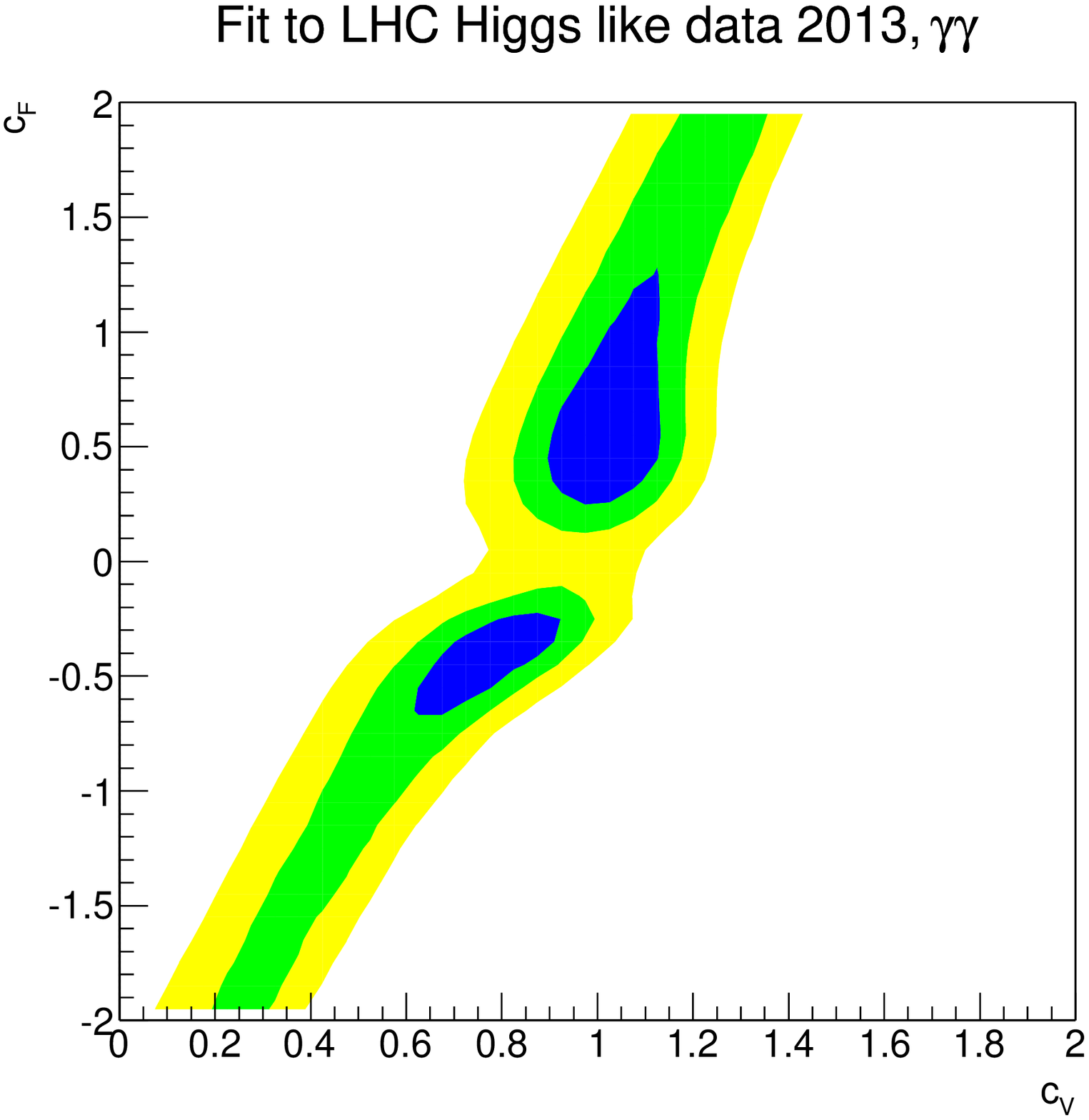}
\begin{center}
{ \it (f)}
\end{center}
\vspace*{-0.0cm}
\end{minipage}

\vspace{10mm}
\caption[]{\label{12_13} (a),(b),(c) - \it $\chi^2$ fits in the $(c_V,c_F)$ plane calculated using post-Moriond 2012 data; (d),(e),(f) - the same fits calculated using LC 2013 data (see Table \ref{lc2013}). (a),(d) - $b \bar b$ and $\tau^+ \tau^-$ channels; (b),(e) - $WW^*$ and $ZZ^*$ channels (including VBF) combined with $\gamma \gamma$ VBF, (c),(f) - $\gamma \gamma$ channels including VBF.}

\end{figure}

Latest data from LC2013 \cite{lc2013} (Hamburg, May 2013) are represented in Table \ref{lc2013}. Significant improvements of the precision have been achieved for $b \bar b$ and $\tau^+ \tau^-$ channels. In the 2012 data the signal strength error both of ATLAS and CMS was about 2, with the three-four times decrease reported the beginning of 2013. Improvements for the $WW$, $ZZ$ and inclusive $\gamma \gamma$ channels have been also quite substantial, reducing the signal strength error approximately by a factor of two. The $H\to W^+ W^-$ signal strength reported by ATLAS was reduced to 1. CMS signal strength for the $\gamma \gamma$ channel was reported on the level of 0.77 in 2013, compared with 1.6 in the earlier data processing. ATLAS reduced it from 1.8 to 1.65. Some improvement for the VBF $\gamma \gamma$ channel was found. A new result for the signal strength 1.1 with the reduced error in the $\tau^+ \tau^-$ channel by CMS improves the primary value of 0.7. The contours generated with post-Moriond 2012 and preliminary LC2013 data again for the three groups of channels (a),(b) and (c), see above, are shown in Fig.\ref{12_13}. As a result, the area at negative $c$ disappeared almost completely while contours of the positive ($a$,$c$) quadrant demonstrate the consistency of the SM hypothesis with the data on the level of 95\%. These modifications are in a qualitative agreement with the global combination from \cite{ellis_you_update} which is based on Moriond 2013 (La Thuile, March 2013) experimental data \cite{moriond2013}.      

\begin{figure}[t]
\begin{center}
\input{WW.tex}
\end{center}
\caption{\label{ww} Signal diagrams for the process
$gg \to WW \to \nu_e {\bar \nu}_e e^+ e^-$}
\end{figure}
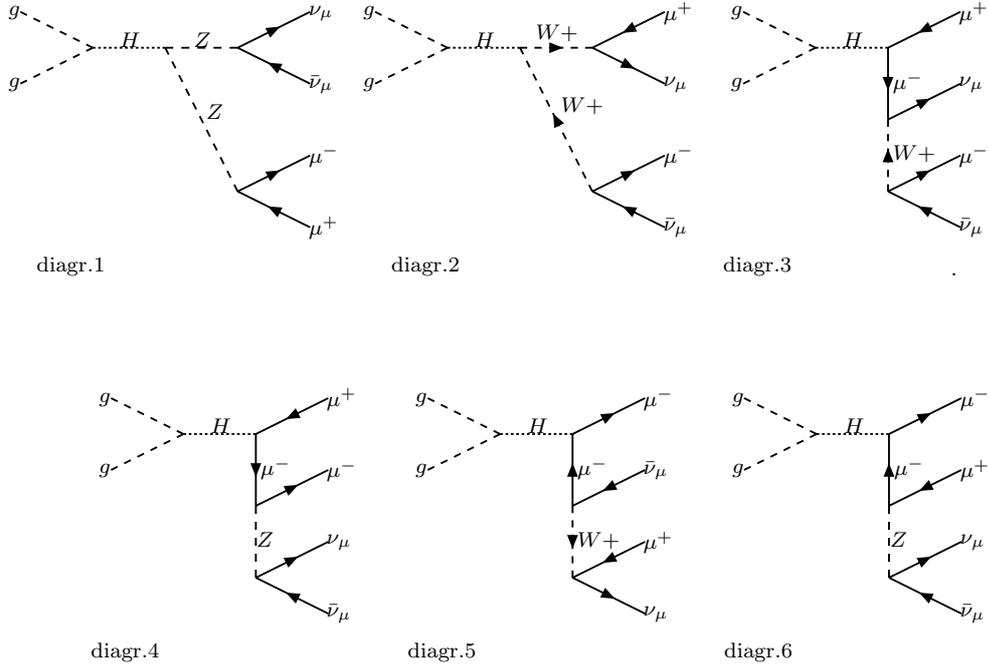

\section{Conclusions}
The LHC data in various channels of Higgs boson production have been analysed in the framework of the Standard Model extension by the dimension-six effective operators.
In order to understand the consistency between the SM consequences and the experimental data a number of global fits for the signal strengths in various channels was performed and the exclusion regions in the ($c_V$,$c_F$) anomalous coupling plane were reconstructed using post-Moriond 2012 data and recent LC 2013 data. In agreement with \cite{fit1} two best fit regions were found for positive and negative values of $c_F$, demonstrating consistency of the SM hypothesis with the post-Moriond 2012 data on the level of 82\%. In the infinitely small width approximation (or {\it production$\times$decay approximation}) we reproduced practically identically the results of \cite{fit1}, although different physics frameworks (effective operator bases) for the rescaling of the Higgs-boson and the Higgs-fermion couplings are used in the analyses.  Evaluations beyond the infinitely small width approximation demonstrate visible departures of the exclusion contours for the combination of $H\to W^+ W^-, ZZ$ and $\gamma \gamma$ channels, however insignificant for the global fits. Improvements of the precision achieved in 2013 \cite{lc2013} excluded practically completely the region with negative values of $c_F$ showing consistency of the SM hypothesis with the 2013 data on the level of 95\%.    

\begin{center}
{\bf Acknowledgements}
\end{center}

The work of E.B., V.B. and M.D. was partially supported by RFBR grant 12-02-93108. M.D. is grateful to Andre David for useful discussion.

\newpage

\newpage


\newpage


\newpage


\end{document}

%% file: WW.tex

{
\unitlength=1.0 pt
\SetScale{1.0}
\SetWidth{0.7}      
\scriptsize    
. \hspace{-2mm}
\begin{picture}(93,102)(0,0)
\DashLine(54.0,85.5)(27.0,99.0){3.0} 
\Text(27.0,99.0)[r]{$g$}
\DashLine(54.0,85.5)(27.0,72.0){3.0} 
\Text(27.0,72.0)[r]{$g$}
\DashLine(54.0,85.5)(81.0,85.5){1.0}
\Text(68.2,86.2)[b]{$H$}
\DashLine(81.0,85.5)(108.0,85.5){3.0} 
\Text(95.2,86.2)[b]{$Z$}
\DashLine(81.0,85.5)(108.0,31.5){3.0} 
\Text(96.8,59.2)[lb]{$Z$}
\ArrowLine(108.0,85.5)(135.0,99.0) 
\Text(135.0,99.0)[l]{$\nu_\mu$}
\ArrowLine(135.0,72.0)(108.0,85.5) 
\Text(135.0,72.0)[l]{${\bar \nu}_\mu$}
\ArrowLine(108.0,31.5)(135.0,45.0) 
\Text(135.0,45.0)[l]{$\mu^-$}
\ArrowLine(135.0,18.0)(108.0,31.5) 
\Text(135.0,18.0)[l]{$\mu^+$}
\Text(46,0)[b] {diagr.1}
\end{picture} 
\vspace{5mm}
{} \qquad\allowbreak
\hspace{5mm}
\begin{picture}(93,102)(0,0)
\DashLine(54.0,85.5)(27.0,99.0){3.0} 
\Text(27.0,99.0)[r]{$g$}
\DashLine(54.0,85.5)(27.0,72.0){3.0} 
\Text(27.0,72.0)[r]{$g$}
\DashLine(54.0,85.5)(81.0,85.5){1.0}
\Text(68.2,86.2)[b]{$H$}
\DashArrowLine(81.0,85.5)(108.0,85.5){3.0} 
\Text(95.2,88.5)[b]{$W+$}
\DashArrowLine(108.0,31.5)(81.0,85.5){3.0} 
\Text(96.8,60.8)[lb]{$W+$}
\ArrowLine(135.0,99.0)(108.0,85.5) 
\Text(135.0,99.0)[l]{$\mu^+$}
\ArrowLine(108.0,85.5)(135.0,72.0) 
\Text(135.0,72.0)[l]{$\nu_\mu$}
\ArrowLine(108.0,31.5)(135.0,45.0) 
\Text(135.0,45.0)[l]{$\mu^-$}
\ArrowLine(135.0,18.0)(108.0,31.5) 
\Text(135.0,18.0)[l]{${\bar \nu}_\mu$}
\Text(46,0)[b] {diagr.2}
\end{picture} \ 
\vspace{5mm}
{} \qquad\allowbreak
\hspace{1mm}
\begin{picture}(93,102)(0,0)
\DashLine(67.5,85.5)(40.5,99.0){3.0} 
\Text(40.5,99.0)[r]{$g$}
\DashLine(67.5,85.5)(40.5,72.0){3.0} 
\Text(40.5,72.0)[r]{$g$}
\DashLine(67.5,85.5)(94.5,85.5){1.0}
\Text(81.8,86.2)[b]{$H$}
\ArrowLine(121.5,99.0)(94.5,85.5) 
\Text(121.5,99.0)[l]{$\mu^+$}
\ArrowLine(94.5,85.5)(94.5,58.5) 
\Text(96.8,72.8)[l]{$\mu^-$}
\ArrowLine(94.5,58.5)(121.5,72.0) 
\Text(121.5,72.0)[l]{$\nu_\mu$}
\DashArrowLine(94.5,31.5)(94.5,58.5){3.0} 
\Text(96.8,45.8)[l]{$W+$}
\ArrowLine(94.5,31.5)(121.5,45.0) 
\Text(121.5,45.0)[l]{$\mu^-$}
\ArrowLine(121.5,18.0)(94.5,31.5) 
\Text(121.5,18.0)[l]{${\bar \nu}_\mu$}
\Text(46,0)[b] {diagr.3}
\end{picture} \ 
\vspace{5mm}
{} \qquad\allowbreak
. \hspace{-14mm}
\begin{picture}(93,102)(0,0)
\DashLine(67.5,85.5)(40.5,99.0){3.0} 
\Text(40.5,99.0)[r]{$g$}
\DashLine(67.5,85.5)(40.5,72.0){3.0} 
\Text(40.5,72.0)[r]{$g$}
\DashLine(67.5,85.5)(94.5,85.5){1.0}
\Text(81.8,86.2)[b]{$H$}
\ArrowLine(121.5,99.0)(94.5,85.5) 
\Text(121.5,99.0)[l]{$\mu^+$}
\ArrowLine(94.5,85.5)(94.5,58.5) 
\Text(96.8,72.8)[l]{$\mu^-$}
\ArrowLine(94.5,58.5)(121.5,72.0) 
\Text(121.5,72.0)[l]{$\mu^-$}
\DashLine(94.5,58.5)(94.5,31.5){3.0} 
\Text(95.2,45.8)[l]{$Z$}
\ArrowLine(94.5,31.5)(121.5,45.0) 
\Text(121.5,45.0)[l]{$\nu_\mu$}
\ArrowLine(121.5,18.0)(94.5,31.5) 
\Text(121.5,18.0)[l]{${\bar \nu}_\mu$}
\Text(46,0)[b] {diagr.4}
\end{picture} \ 
\vspace{5mm}
{} \qquad\allowbreak
\begin{picture}(93,102)(0,0)
\DashLine(67.5,85.5)(40.5,99.0){3.0} 
\Text(40.5,99.0)[r]{$g$}
\DashLine(67.5,85.5)(40.5,72.0){3.0} 
\Text(40.5,72.0)[r]{$g$}
\DashLine(67.5,85.5)(94.5,85.5){1.0}
\Text(81.8,86.2)[b]{$H$}
\ArrowLine(94.5,85.5)(121.5,99.0) 
\Text(121.5,99.0)[l]{$\mu^-$}
\ArrowLine(94.5,58.5)(94.5,85.5) 
\Text(96.8,72.8)[l]{$\mu^-$}
\ArrowLine(121.5,72.0)(94.5,58.5) 
\Text(121.5,72.0)[l]{${\bar \nu}_\mu$}
\DashArrowLine(94.5,58.5)(94.5,31.5){3.0} 
\Text(96.8,45.8)[l]{$W+$}
\ArrowLine(121.5,45.0)(94.5,31.5) 
\Text(121.5,45.0)[l]{$\mu^+$}
\ArrowLine(94.5,31.5)(121.5,18.0) 
\Text(121.5,18.0)[l]{$\nu_\mu$}
\Text(46,0)[b] {diagr.5}
\end{picture} \ 
\vspace{5mm}
{} \qquad\allowbreak
\begin{picture}(93,102)(0,0)
\DashLine(67.5,85.5)(40.5,99.0){3.0} 
\Text(40.5,99.0)[r]{$g$}
\DashLine(67.5,85.5)(40.5,72.0){3.0} 
\Text(40.5,72.0)[r]{$g$}
\DashLine(67.5,85.5)(94.5,85.5){1.0}
\Text(81.8,86.2)[b]{$H$}
\ArrowLine(94.5,85.5)(121.5,99.0) 
\Text(121.5,99.0)[l]{$\mu^-$}
\ArrowLine(94.5,58.5)(94.5,85.5) 
\Text(96.8,72.8)[l]{$\mu^-$}
\ArrowLine(121.5,72.0)(94.5,58.5) 
\Text(121.5,72.0)[l]{$\mu^+$}
\DashLine(94.5,58.5)(94.5,31.5){3.0} 
\Text(95.2,45.8)[l]{$Z$}
\ArrowLine(94.5,31.5)(121.5,45.0) 
\Text(121.5,45.0)[l]{$\nu_\mu$}
\ArrowLine(121.5,18.0)(94.5,31.5) 
\Text(121.5,18.0)[l]{${\bar \nu}_\mu$}
\Text(46,0)[b] {diagr.6}
\end{picture} 
}

%% file: ZZ.tex
{
\unitlength=1.0 pt
\SetScale{1.0}
\SetWidth{0.7}      
\scriptsize    
\hspace{-12mm}
\begin{picture}(93,102)(0,0)
\DashLine(54.0,85.5)(27.0,99.0){3.0} 
\Text(27.0,99.0)[r]{$G$}
\DashLine(54.0,85.5)(27.0,72.0){3.0} 
\Text(27.0,72.0)[r]{$G$}
\DashLine(54.0,85.5)(81.0,85.5){1.0}
\Text(68.2,86.2)[b]{$H$}
\DashLine(81.0,85.5)(108.0,85.5){3.0} 
\Text(95.2,86.2)[b]{$A$}
\DashLine(81.0,85.5)(108.0,31.5){3.0} 
\Text(96.8,59.2)[lb]{$A$}
\ArrowLine(108.0,85.5)(135.0,99.0) 
\Text(135.0,99.0)[l]{$\mu^-$}
\ArrowLine(135.0,72.0)(108.0,85.5) 
\Text(135.0,72.0)[l]{$\mu^+$}
\ArrowLine(108.0,31.5)(135.0,45.0) 
\Text(135.0,45.0)[l]{$\mu^-$}
\ArrowLine(135.0,18.0)(108.0,31.5) 
\Text(135.0,18.0)[l]{$\mu^+$}
\Text(46,0)[b] {diagr.1}
\end{picture} \ 
\vspace{5mm}
{} \qquad\allowbreak
\hspace{5mm}
\begin{picture}(93,102)(0,0)
\DashLine(54.0,85.5)(27.0,99.0){3.0} 
\Text(27.0,99.0)[r]{$G$}
\DashLine(54.0,85.5)(27.0,72.0){3.0} 
\Text(27.0,72.0)[r]{$G$}
\DashLine(54.0,85.5)(81.0,85.5){1.0}
\Text(68.2,86.2)[b]{$H$}
\DashLine(81.0,85.5)(108.0,85.5){3.0} 
\Text(95.2,86.2)[b]{$Z$}
\DashLine(81.0,85.5)(108.0,31.5){3.0} 
\Text(96.8,59.2)[lb]{$Z$}
\ArrowLine(108.0,85.5)(135.0,99.0) 
\Text(135.0,99.0)[l]{$\mu^-$}
\ArrowLine(135.0,72.0)(108.0,85.5) 
\Text(135.0,72.0)[l]{$\mu^+$}
\ArrowLine(108.0,31.5)(135.0,45.0) 
\Text(135.0,45.0)[l]{$\mu^-$}
\ArrowLine(135.0,18.0)(108.0,31.5) 
\Text(135.0,18.0)[l]{$\mu^+$}
\Text(46,0)[b] {diagr.2}
\end{picture} \ 
\vspace{5mm}
{} \qquad\allowbreak
\begin{picture}(93,102)(0,0)
\DashLine(67.5,85.5)(40.5,99.0){3.0} 
\Text(40.5,99.0)[r]{$G$}
\DashLine(67.5,85.5)(40.5,72.0){3.0} 
\Text(40.5,72.0)[r]{$G$}
\DashLine(67.5,85.5)(94.5,85.5){1.0}
\Text(81.8,86.2)[b]{$H$}
\ArrowLine(121.5,99.0)(94.5,85.5) 
\Text(121.5,99.0)[l]{$\mu^+$}
\ArrowLine(94.5,85.5)(94.5,58.5) 
\Text(96.8,72.8)[l]{$\mu^-$}
\ArrowLine(94.5,58.5)(121.5,72.0) 
\Text(121.5,72.0)[l]{$\mu^-$}
\DashLine(94.5,58.5)(94.5,31.5){3.0} 
\Text(95.2,45.8)[l]{$A$}
\ArrowLine(94.5,31.5)(121.5,45.0) 
\Text(121.5,45.0)[l]{$\mu^-$}
\ArrowLine(121.5,18.0)(94.5,31.5) 
\Text(121.5,18.0)[l]{$\mu^+$}
\Text(46,0)[b] {diagr.3}
\end{picture} \ 
\vspace{5mm}
{} \qquad\allowbreak
\begin{picture}(93,102)(0,0)
\DashLine(67.5,85.5)(40.5,99.0){3.0} 
\Text(40.5,99.0)[r]{$G$}
\DashLine(67.5,85.5)(40.5,72.0){3.0} 
\Text(40.5,72.0)[r]{$G$}
\DashLine(67.5,85.5)(94.5,85.5){1.0}
\Text(81.8,86.2)[b]{$H$}
\ArrowLine(121.5,99.0)(94.5,85.5) 
\Text(121.5,99.0)[l]{$\mu^+$}
\ArrowLine(94.5,85.5)(94.5,58.5) 
\Text(96.8,72.8)[l]{$\mu^-$}
\ArrowLine(94.5,58.5)(121.5,72.0) 
\Text(121.5,72.0)[l]{$\mu^-$}
\DashLine(94.5,58.5)(94.5,31.5){3.0} 
\Text(95.2,45.8)[l]{$Z$}
\ArrowLine(94.5,31.5)(121.5,45.0) 
\Text(121.5,45.0)[l]{$\mu^-$}
\ArrowLine(121.5,18.0)(94.5,31.5) 
\Text(121.5,18.0)[l]{$\mu^+$}
\Text(46,0)[b] {diagr.4}
\end{picture} \ 
\vspace{5mm}
{} \qquad\allowbreak
\begin{picture}(93,102)(0,0)
\DashLine(67.5,85.5)(40.5,99.0){3.0} 
\Text(40.5,99.0)[r]{$G$}
\DashLine(67.5,85.5)(40.5,72.0){3.0} 
\Text(40.5,72.0)[r]{$G$}
\DashLine(67.5,85.5)(94.5,85.5){1.0}
\Text(81.8,86.2)[b]{$H$}
\ArrowLine(121.5,99.0)(94.5,85.5) 
\Text(121.5,99.0)[l]{$\mu^+$}
\ArrowLine(94.5,85.5)(94.5,58.5) 
\Text(96.8,72.8)[l]{$\mu^-$}
\ArrowLine(94.5,58.5)(121.5,72.0) 
\Text(121.5,72.0)[l]{$\mu^-$}
\DashLine(94.5,58.5)(94.5,31.5){1.0}
\Text(95.2,45.8)[l]{$H$}
\ArrowLine(94.5,31.5)(121.5,45.0) 
\Text(121.5,45.0)[l]{$\mu^-$}
\ArrowLine(121.5,18.0)(94.5,31.5) 
\Text(121.5,18.0)[l]{$\mu^+$}
\Text(46,0)[b] {diagr.5}
\end{picture} \ 
\vspace{5mm}
{} \qquad\allowbreak
\begin{picture}(93,102)(0,0)
\DashLine(67.5,85.5)(40.5,99.0){3.0} 
\Text(40.5,99.0)[r]{$G$}
\DashLine(67.5,85.5)(40.5,72.0){3.0} 
\Text(40.5,72.0)[r]{$G$}
\DashLine(67.5,85.5)(94.5,85.5){1.0}
\Text(81.8,86.2)[b]{$H$}
\ArrowLine(94.5,85.5)(121.5,99.0) 
\Text(121.5,99.0)[l]{$\mu^-$}
\ArrowLine(94.5,58.5)(94.5,85.5) 
\Text(96.8,72.8)[l]{$\mu^-$}
\ArrowLine(121.5,72.0)(94.5,58.5) 
\Text(121.5,72.0)[l]{$\mu^+$}
\DashLine(94.5,58.5)(94.5,31.5){3.0} 
\Text(95.2,45.8)[l]{$A$}
\ArrowLine(94.5,31.5)(121.5,45.0) 
\Text(121.5,45.0)[l]{$\mu^-$}
\ArrowLine(121.5,18.0)(94.5,31.5) 
\Text(121.5,18.0)[l]{$\mu^+$}
\Text(46,0)[b] {diagr.6}
\end{picture} \ 
\vspace{5mm}
{} \qquad\allowbreak
\begin{picture}(93,102)(0,0)
\DashLine(67.5,85.5)(40.5,99.0){3.0} 
\Text(40.5,99.0)[r]{$G$}
\DashLine(67.5,85.5)(40.5,72.0){3.0} 
\Text(40.5,72.0)[r]{$G$}
\DashLine(67.5,85.5)(94.5,85.5){1.0}
\Text(81.8,86.2)[b]{$H$}
\ArrowLine(94.5,85.5)(121.5,99.0) 
\Text(121.5,99.0)[l]{$\mu^-$}
\ArrowLine(94.5,58.5)(94.5,85.5) 
\Text(96.8,72.8)[l]{$\mu^-$}
\ArrowLine(121.5,72.0)(94.5,58.5) 
\Text(121.5,72.0)[l]{$\mu^+$}
\DashLine(94.5,58.5)(94.5,31.5){3.0} 
\Text(95.2,45.8)[l]{$Z$}
\ArrowLine(94.5,31.5)(121.5,45.0) 
\Text(121.5,45.0)[l]{$\mu^-$}
\ArrowLine(121.5,18.0)(94.5,31.5) 
\Text(121.5,18.0)[l]{$\mu^+$}
\Text(46,0)[b] {diagr.7}
\end{picture} \ 
\vspace{5mm}
{} \qquad\allowbreak
\begin{picture}(93,102)(0,0)
\DashLine(67.5,85.5)(40.5,99.0){3.0} 
\Text(40.5,99.0)[r]{$G$}
\DashLine(67.5,85.5)(40.5,72.0){3.0} 
\Text(40.5,72.0)[r]{$G$}
\DashLine(67.5,85.5)(94.5,85.5){1.0}
\Text(81.8,86.2)[b]{$H$}
\ArrowLine(94.5,85.5)(121.5,99.0) 
\Text(121.5,99.0)[l]{$\mu^-$}
\ArrowLine(94.5,58.5)(94.5,85.5) 
\Text(96.8,72.8)[l]{$\mu^-$}
\ArrowLine(121.5,72.0)(94.5,58.5) 
\Text(121.5,72.0)[l]{$\mu^+$}
\DashLine(94.5,58.5)(94.5,31.5){1.0}
\Text(95.2,45.8)[l]{$H$}
\ArrowLine(94.5,31.5)(121.5,45.0) 
\Text(121.5,45.0)[l]{$\mu^-$}
\ArrowLine(121.5,18.0)(94.5,31.5) 
\Text(121.5,18.0)[l]{$\mu^+$}
\Text(46,0)[b] {diagr.8}
\end{picture} \ 
\vspace{5mm}
{} \qquad\allowbreak
\begin{picture}(93,102)(0,0)
\DashLine(54.0,85.5)(27.0,99.0){3.0} 
\Text(27.0,99.0)[r]{$G$}
\DashLine(54.0,85.5)(27.0,72.0){3.0} 
\Text(27.0,72.0)[r]{$G$}
\DashLine(54.0,85.5)(81.0,85.5){1.0}
\Text(68.2,86.2)[b]{$H$}
\DashLine(81.0,85.5)(108.0,85.5){1.0}
\Text(95.2,86.2)[b]{$H$}
\DashLine(81.0,85.5)(108.0,31.5){1.0}
\Text(96.8,59.2)[lb]{$H$}
\ArrowLine(108.0,85.5)(135.0,99.0) 
\Text(135.0,99.0)[l]{$\mu^-$}
\ArrowLine(135.0,72.0)(108.0,85.5) 
\Text(135.0,72.0)[l]{$\mu^+$}
\ArrowLine(108.0,31.5)(135.0,45.0) 
\Text(135.0,45.0)[l]{$\mu^-$}
\ArrowLine(135.0,18.0)(108.0,31.5) 
\Text(135.0,18.0)[l]{$\mu^+$}
\Text(46,0)[b] {diagr.9}
\end{picture} \ 
\vspace{5mm}
{} \qquad\allowbreak
\begin{picture}(93,102)(0,0)
\DashLine(67.5,85.5)(40.5,85.5){3.0} 
\Text(40.5,85.5)[r]{$G$}
\DashLine(67.5,85.5)(67.5,31.5){3.0} 
\Text(68.2,59.2)[l]{$G$}
\DashLine(67.5,85.5)(94.5,85.5){1.0}
\Text(81.8,86.2)[b]{$H$}
\DashLine(67.5,31.5)(40.5,31.5){3.0} 
\Text(40.5,31.5)[r]{$G$}
\DashLine(67.5,31.5)(94.5,31.5){1.0}
\Text(81.8,32.2)[b]{$H$}
\ArrowLine(94.5,31.5)(121.5,45.0) 
\Text(121.5,45.0)[l]{$\mu^-$}
\ArrowLine(121.5,18.0)(94.5,31.5) 
\Text(121.5,18.0)[l]{$\mu^+$}
\ArrowLine(94.5,85.5)(121.5,99.0) 
\Text(121.5,99.0)[l]{$\mu^-$}
\ArrowLine(121.5,72.0)(94.5,85.5) 
\Text(121.5,72.0)[l]{$\mu^+$}
\Text(46,0)[b] {diagr.10}
\end{picture} \ 
\vspace{5mm}
}

%% file: dim6.bbl
\begin{thebibliography}{99}

\bibitem{signal}
G. Aad {\it et al.}(ATLAS Collaboration), {\it Observation of a new particle in the search for the SM Higgs boson with the ATLAS detector at the LHC}, Phys. Lett. {\bf B716}, 1 (2012) (arXiv:1207.7214[hep-ex])\\
S. Chatrchyan {\it et al.}(CMS Collaboration), {\it Observation of a new boson at a mass of 125 GeV with the CMS experiment at the LHC}, Phys. Lett. {\bf B716}, 30 (2012) (arXiv:1207.7235[hep-ex])

\bibitem{first}
E. Eichten, K. Lane, M. Peskin, {\it New tests for quark and lepton substructure}, Phys. Rev. Lett. {\bf 50}, 811 (1983)\\
H. Weldon, A.Zee, {\it Opearator analysis of new physics}, Nucl. Phys. {\bf B173}, 269 (1980)

\bibitem{following}
K.~Whisnant, J.M.~Yang, B.-L.~Young, X.~Zhang, {\it Dimension six CP-conserving operators of the third family quarks and their effects on collider observables}, Phys. Rev. {\bf D56} (1997) 467 (hep-ph/9702305)\\
S. Dimopoulos, J. Ellis, {\it Challenges for Extended Technicolor Theories}, Nucl. Phys. {\bf B182}, 505 (1981)\\
R. Cahn, H. Harari, {\it Bounds on the Masses of Neutral Generation Changing Gauge Bosons}, Nucl. Phys. {\bf B176}, 135 (1980)\\
Y. Maehara, T. Yanagida, {\it Gauge symmetry of horisontal flavor}, Progr. Theor. Phys. {\bf 61}, 1434 (1979)

\bibitem{buchmuller}
W.~Buchmuller, D.~Wyler, {\it Effective Lagrangian analysis of new interactins and flavour conservation}, Nucl.Phys. {\bf B268}, 621 (1986)
 
\bibitem{grzadkowski}
B. Grzadkowski, M. Iskrzynski, M. Misiak, J. Rosiek, 	
{\it Dimension-Six Terms in the Standard Model Lagrangian},
JHEP {\bf 1010}, 085 (2010) (arXiv:1008.4884 [hep-ph])

\bibitem{contino_basis}
R. Contino, M. Ghezzi, C. Grojean, M. Muhlleitner, M. Spira, {\it 	
Effective Lagrangian for a light Higgs-like scalar},
JHEP {\bf 1307}, 035 (2013) (arXiv:1303.3876 [hep-ph])

\bibitem{passarino}
G. Passarino, {\it NLO inspired effective Lagrangians for Higgs physics}, Nucl.Phys. {\bf B868}, 416 (2013) (arXiv:1209.5538 [hep-ph])

\bibitem{eboli} 
T. Corbett, O.J.P. Eboli, J. Gonzalez-Fraile, M.C. Gonzalez-Garcia, 	
{\it Constraining anomalous Higgs interactions},
Phys.Rev. {\bf D86}, 075013 (2012) (arXiv:1207.1344 [hep-ph])

\bibitem{tensor}
B. Dumont, S. Fichet, G. von Gersdorff, {\it	
A Bayesian view of the Higgs sector with higher dimensional operators},
JHEP {\bf 1307}, 065 (2013) (arXiv:1304.3369 [hep-ph]) 

\bibitem{distributions}
S. Banerjee, S. Mukhopadhyay, B. Mukhopadhyaya,
{\it Higher dimensional operators and LHC Higgs data : the role of modified kinematics},
arXiv:1308.4860 [hep-ph], 2013

\bibitem{fullset}
We-Fu Chang, Wei-Ping Pan, Fanrong Xu, {\it	
An effective gauge-Higgs operators analysis of new physics associated with the Higgs},
Phys.Rev. {\bf D88}, 033004 (2013) (arXiv:1303.7035 [hep-ph])

Kingman Cheung, J. S. Lee, Po-Yan Tseng,
{\it Higgs Precision (Higgcision) Era begins}, arXiv:1302.3794 [hep-ph], 2013

G. Belanger, B. Dumont, U. Ellwanger, J.F. Gunion, S. Kraml,
{\it Higgs Couplings at the End of 2012}, JHEP {\bf 1302}, 053 (2013) (arXiv:1212.5244 [hep-ph]) 

A. Azatov, J. Galloway, 	
{\it Electroweak Symmetry Breaking and the Higgs Boson: Confronting Theories at Colliders},
Int.J.Mod.Phys. {\bf A28} 1330004 (2013) (arXiv:1212.1380 [hep-ph])

T. Corbett, O.J.P. Eboli, J. Gonzalez-Fraile, M.C. Gonzalez-Garcia,
{\it Robust Determination of the Higgs Couplings: Power to the Data}
Phys.Rev. {\bf D87}, 015022 (2013) (arXiv:1211.4580 [hep-ph])

E. Masso, V. Sanz,
{\it Limits on Anomalous Couplings of the Higgs to Electroweak Gauge Bosons from LEP and LHC},
Phys.Rev. {\bf D87}, 033001 (2013) (arXiv:1211.1320 [hep-ph])

S. Dittmaier, M. Schumacher, {\it	
The Higgs Boson in the Standard Model - From LEP to LHC: Expectations, Searches, and Discovery of a Candidate}, Prog.Part.Nucl.Phys. {\bf 70}, 1-54, 2013 (arXiv:1211.4828 [hep-ph])

G. Cacciapaglia, A. Deandrea, G. Drieu La Rochelle, J.-B. Flament,
{\it Higgs couplings beyond the Standard Model}
JHEP {\bf 1303}, 029 (2013) (arXiv:1210.8120 [hep-ph])

B. Dobrescu, J. Lykken, {\it Coupling spans of the Higgs-like boson},  JHEP {\bf 1302}, 073 (2013) (arXiv:1210.3342 [hep-ph])

B. Batell, S. Gori, Lian-Tao Wang,
{\it Higgs Couplings and Precision Electroweak Data},
JHEP {\bf 1301}, 139 (2013) (arXiv:1209.6382 [hep-ph])

A. Djouadi, 	
{\it Precision Higgs coupling measurements at the LHC through ratios of production cross sections}, arXiv:1208.3436 [hep-ph], 2012

T. Plehn, M. Rauch,
{\it Higgs Couplings after the Discovery}, 
Europhys.Lett. {\bf 100}, 11020 (2012) (arXiv:1207.6108 [hep-ph])

F. Bonnet, T. Ota, M. Rauch, W. Winter,
{\it Interpretation of precision tests in the Higgs sector in terms of physics beyond the Standard Model}, Phys.Rev. {\it D86}, 093014 (2012) (arXiv:1207.4599 [hep-ph])

S. Banerjee, S. Mukhopadhyay, B. Mukhopadhyaya,
{\it New Higgs interactions and recent data from the LHC and the Tevatron},
JHEP {\bf 1210}, 062 (2012) (arXiv:1207.3588 [hep-ph])

D. Carmi, A. Falkowski, E. Kuflik, T. Volansky, J. Zupan,
{\it Higgs After the Discovery: A Status Report}
JHEP {\bf 1210}, 196 (2012) (arXiv:1207.1718 [hep-ph])

M. Montull, F. Riva	
{\it Higgs discovery: the beginning or the end of natural EWSB?}
JHEP {\bf 1211}, 018 (2012) 018 (arXiv:1207.1716 [hep-ph])

P. Giardino, K. Kannike, M. Raidal, A. Strumia,
{\it Is the resonance at 125 GeV the Higgs boson?}
Phys.Lett. {\bf B718}, 469 (2012) (arXiv:1207.1347 [hep-ph])

I. Low, J. Lykken, G. Shaughnessy,	
{\it Have We Observed the Higgs (Imposter)?}
Phys.Rev. {\bf D86}, 093012 (2012) 093012 (arXiv:1207.1093 [hep-ph])

M. Klute, R. Lafaye, T. Plehn, M. Rauch, D. Zerwas,
{\it Measuring Higgs Couplings from LHC Data}, Phys.Rev.Lett. {\bf 109}, 101801 (2012)
(arXiv:1205.2699 [hep-ph])

A. Azatov, R. Contino, D. Del Re, J. Galloway, M. Grassi, S. Rahatlou, 
{\it Determining Higgs couplings with a model-independent analysis of h ->gamma gamma},
JHEP {\bf 1206}, 134 (2012) 134 (arXiv:1204.4817 [hep-ph])

Tianjin Li, Xia Wan, You-kai Wang, Shou-hua Zhu, {\it
Constraints on the Universal Varying Yukawa Couplings: from SM-like to Fermiophobic},
JHEP {\bf 1209} (2012) 086 (arXiv:1203.5083 [hep-ph])

P. Giardino, K. Kannike, M. Raidal, A. Strumia, {\it 	
Reconstructing Higgs boson properties from the LHC and Tevatron data},
JHEP {\bf 1206}, 117 (2012) (arXiv:1203.4254 [hep-ph])

D. Carmi, A. Falkowski, E. Kuflik, T. Volansky, {\it	
Interpreting LHC Higgs Results from Natural New Physics Perspective},
JHEP {\bf 1207}, 136 (2012) (arXiv:1202.3144 [hep-ph])

\bibitem{ellis_you}
J. Ellis, T. You, {\it Global Analysis of the Higgs Candidate with Mass ~ 125 GeV }, JHEP {\bf 1209}, 123 (2012) (arXiv:1207.1693 [hep-ph])

\bibitem{ellis_you_0}
J. Ellis, T. You, 	
{\it Global Analysis of Experimental Constraints on a Possible Higgs-Like Particle with Mass ~ 125 GeV}, JHEP {\bf 1206}, 140 (2012) (arXiv:1204.0464 [hep-ph])

\bibitem{silh}
G.F. Guidice, C. Grojean, A. Pomarol, R. Ratazzi, {\it The strongly interacting light Higgs}, JHEP {\bf 0706}, 045 (2007)(arXiv:hep-ph/0703164)

\bibitem{silh_lgrng}
R. Grober, M. Muhlleitner, {\it Composite Higgs boson pair production at the LHC}, JHEP {\bf 1106}, 020 (2011) (arXiv:1012.1562 [hep-ph])

R. Contino, C. Grojean, M. Moretti, F. Piccinini, R. Ratazzi, {\it Strong Double Higgs Production at the LHC} JHEP {\bf 1005}, 089 (2010) (arXiv:1002.1011 [hep-ph])

\bibitem{silheffects}
M. Dolan, C. Englert, M. Spannowsky, {\it New physics in LHC Higgs boson pair-production}, arXiv:1210.8166 [hep-ph], 2012

A. Azatov, R. Contino, J. Galloway, {\it Model-Independent Bounds on a Light Higgs}, JHEP {\bf 1204}, 127 (2012) (arXiv:1202.3144 [hep-ph])

\bibitem{fit1}
J.R. Espinosa, C. Grojean, M. Muhlleitner, M. Trott, 
{\it Fingerprinting Higgs Suspects at the LHC}, JHEP {\bf 1205}, 097 (2012) (arXiv:1202.3697 [hep-ph]) 
 
J.R. Espinosa, C. Grojean, M. Muhlleitner, M. Trott, {\it First glimpses at Higgs' face}, JHEP {\bf 1212}, 045 (2012) (arXiv:1207.1717 [hep-ph])

\bibitem{cms_pas}
CMS Physics Analysis Summaries, {\it Combination of standard model Higgs boson searches and measurements of the properties of the new boson with a mass near 125 GeV}, CMS-PAS-HIG-13-005,
see {\tt https://cds.cern.ch/record/1542387}

ATLAS Collaboration, {\it Measurements of Higgs boson production and couplings in diboson final states with the ATLAS detector at the LHC}, arXiv:1307.1427v1 [hep-ex]

\bibitem{hxswg}
A.David, A.Denner, M.Duhrssen, M.Grazzini, C.Grojean, G.Passarino, M.Schumacher, M.Spira, G.Weiglein, M.Zanetti, {\it LHC HXSWG interim recommendations to explore the coupling structure of a Higgs-like particle}, arXiv:1209.0040 [hep-ph], 2012

\bibitem{moriond}
F.Hubaut(ATLAS) and G.Gomez-Ceballos(CMS), presentations at Rencontres de Moriond 2013, see {\tt \protect{http://}moriond.in2p3.fr}

\bibitem{05djouadi}
A. Djouadi, {\it 	
The Anatomy of electro-weak symmetry breaking. I: The Higgs boson in the standard model}, Phys.Rept. {\bf 457}, 1 (2008) 1(hep-ph/0503172)

\bibitem{76ellis}
J. Ellis, M.K. Gaillard, D.V. Nanopoulos, {\it A phenomenological profile of the Higgs boson}, Nucl.Phys. {\bf B106}, 292 (1976)

\bibitem{hdecay}
A.~Djouadi, J.~Kalinowski and M.~Spira, Comput.Phys.Commun, {\bf 108},
56 (1998) (hep-ph/9704448)



\bibitem{fit2}
J. R. Espinosa, M. Muhlleitner, C. Grojean, M. Trott, {\it  Probing for Invisible Higgs Decays with Global Fits}, JHEP {\bf 1209}, 126 (2012) (arXiv:1205.6790 [hep-ph]) 

S. Kraml, B.C. Allanach, M. Mangano, H.B. Prosper, S. Sekmen, C. Balazs, A. Barr, P. Bechtle, G. Belanger, A. Belyaev et al., {\it Searches for New Physics: Les Houches Recommendations for the Presentation of LHC Results}, Eur.Phys.J. {\bf C72}, 1976 (2012) (arXiv:1203.2489 [hep-ph])

A. Azatov, R. Contino, J. Galloway, {\it Model-Independent Bounds on a Light Higgs}, JHEP {\bf 1204}, 127 (2012) 127, Erratum-ibid. {\bf 1304}, 140 (2013) (arXiv:1202.3415 [hep-ph])

\bibitem{lhchwg}
S. Dittmaier et al., {\it LHC Higgs Cross Section Working Group}, hep-ph/1101.0593, 2011

\bibitem{LEP2}
M.Grunewald, G.Passarino et.al., {\it Four fermion production in electron-positron collisions},
Reports of the working groups on pecision calculations
for LEP2 physics, CERN report 2000-09-A, p. 1-135 (hep-ph/0005309)


\bibitem{CompHEP}
E. Boos {\it et al.},  {\it CompHEP 4.4: automatic computations from Lagrangians to events}, Nucl. Instrum. Meth. {\bf A534}, 250 (2004) (hep-ph/0403113)
A.~Pukhov {\it et al.}, {\it CompHEP 3.3 users' manual}, hep-ph/9908288; \\
see also {\tt http://theory.npi.msu.su/comphep}

\bibitem{lc2013}
D. Zanzi (on behalf of ATLAS collaboration), {\it Latest Higgs results from ATLAS},
A. Savin (on behalf of CMS collaboration), {\it Higgs results from CMS},
reported on the European Linear Collider Workshop, 27 - 31 May 2013, DESY, Hamburg.

\bibitem{ellis_you_update}
J. Ellis, T. You, {\it Updated global analysis of Higgs couplings}, arXiv:1303.3879 [hep-ph], 2013

\bibitem{moriond2013}
Rencontres de Moriond 2013, Higgs, QCD and Electroweak sessions, see {\tt http:\protect{//}moriond.in2p3.fr/QCD/2013/MorQCD13Prog.html}


\end{thebibliography}
